\documentclass[prl, reprint, preprintnumbers, longbibliography]{revtex4-2}

\usepackage{amsfonts}
\usepackage{amssymb}
\usepackage{amsmath}
\usepackage{graphicx}
\usepackage{dcolumn}
\usepackage{bm}
\usepackage{color}
\usepackage{braket}
\usepackage[breaklinks, hidelinks]{hyperref}
\usepackage{float}
\usepackage[utf8]{inputenc}

\begin{document}
	
	\title{Fates of Exact Finite-$N$ First-Order Transitions: Coherent Suppression versus Spectral-Collapse Termination}
	
	\author{Jiong Li$^1$}
	\author{Li-Wei Duan$^2$}
	\author{Qing-Hu Chen$^{3,1,}$}
	\email{qhchen@zju.edu.cn}
	\author{Hai-Qing Lin$^{1,}$}
	\email{hqlin@zju.edu.cn}
	
	\affiliation{
		$^1$ Institute for Advanced Study in Physics and School of Physics, Zhejiang University, Hangzhou 310058, China \\
		$^2$ Department of Physics, Zhejiang Normal University, Jinhua 321004, China \\
		$^3$ Zhejiang Key Laboratory of Micro-Nano Quantum Chips and Quantum Control, School of Physics, Zhejiang University, Hangzhou 310058, China 
	}
	
	\date{\today}
	
	\begin{abstract}
		Symmetry-protected ground-state crossings can produce exact first-order transitions (FOTs) even at finite $N$, in contrast to continuous transitions governed by soft-mode criticality. We establish exact Juddian ground-state crossings for arbitrary finite $N$ in the one- and two-photon anisotropic Dicke models and identify a common two-channel structure: the eigenvalues of the coupling matrix set the local normal-state instability thresholds, whereas its determinant fixes the exact crossing. Despite this common structure, the resulting FOTs have distinct fates determined by the bosonic response. In the one-photon model, the overlap between coherent Juddian branches vanishes in the thermodynamic or infinite frequency-ratio limits, asymptotically suppressing the FOT cusp. In the two-photon model, squeezed Juddian states lose normalizability at spectral collapse, thereby terminating the physical FOT line at finite $N$. Although the cusp strength tends to zero as the endpoint is approached, the terminating FOT retains a collective spectral signature: the different-parity gap is predicted to close with an $N$-dependent exponent $\mu_N=(2N+3)/4$, one greater than the cusp termination exponent. By contrast, the same-parity soft mode retains $N$-independent square-root scaling and governs the critical response. Thus spectral collapse can terminate an exact finite-$N$ FOT while preserving its collective imprint in the parity-resolved spectrum.
	\end{abstract}
	
	\maketitle
	
	\paragraph{Introduction.---}
	Continuous quantum phase transitions (QPTs) typically develop ground-state nonanalyticities in the thermodynamic limit or another singular limit~\cite{sachdevQuantumPhaseTransitions2011, vojtaQuantumPhaseTransitions}. Quantum light-matter systems provide paradigmatic examples: the superradiant transition of the Dicke model (DM) in the thermodynamic limit~\cite{dickeCoherenceSpontaneousRadiation1954a, heppEquilibriumStatisticalMechanics1973, wangPhaseTransitionDicke1973, emaryQuantumChaosTriggered2003,  kirtonIntroductionDickeModel2019}, explored in cavity platforms~\cite{baumannDickeQuantumPhase2010, nagyDickemodelPhaseTransition2010, zhangDickemodelSimulationCavityassisted2018}, and finite-component criticality in the quantum Rabi model (QRM) in the infinite frequency-ratio limit~\cite{ashhabSuperradianceTransitionSystem2013, hwangQuantumPhaseTransition2015,  felicettiUniversalSpectralFeatures2020}. These developments have also motivated critical quantum metrology and quantum simulation~\cite{garbeCriticalQuantumMetrology2020a, caiObservationQuantumPhase2021,  chenExperimentalQuantumSimulation2021a}. In these examples, continuous criticality is associated with a bosonic soft mode whose excitation energy vanishes at the critical point.
	
	Ground-state nonanalyticities can also arise at finite size through a distinct mechanism: exact ground-state crossings with unequal energy slopes produce first-order transitions (FOTs)~\cite{ariasQuantumPhaseTransitions2003, wuQuantumPhaseTransitions2004}. Juddian solutions, which yield exact eigenstates at special parameter values, can support symmetry-protected level crossings in quantum-optical models~\cite{juddExactSolutionsClass1979a, emaryExactIsolatedSolutions2002, braakIntegrabilityRabiModel2011, chenExactSolvabilityQuantum2012}. In anisotropic QRMs, opposite-parity Juddian ground-state crossings can generate FOTs~\cite{xieAnisotropicRabiModel2014, chenMultipleGroundstateInstabilities2021}, while related exceptional solutions occur in the excited spectrum of finite-$N$ DMs~\cite{heExactSolvabilityNonintegrability2015}. Ground-state crossings therefore provide a finite-size route to ground-state nonanalyticity without requiring the soft-mode mechanism of continuous criticality.
	
	Anisotropic light-matter models offer a natural setting for investigating how exact finite-N FOTs evolve under different bosonic responses. In anisotropic QRMs and DMs, rotating and counter-rotating interactions define two effective coupling channels~\cite{ buijsmanNonergodicityAnisotropicDicke2017, liuUniversalScalingCritical2017, liNonlinearDynamicsDissipative2022, dasPhaseTransitionsAnisotropic2023, heHierarchicSuperradiantPhases2026a}. Their one- and two-photon realizations share this coupling structure but differ in their bosonic responses: the former involves coherent displacement, whereas the latter involves squeezing and can exhibit spectral collapse at finite coupling strength. Two-photon interactions can be engineered in trapped-ion~\cite{felicettiSpectralCollapseTwophonon2015, casanovaConnectingNthOrder2018, pueblaQuantumSimulationMultiphoton2019} and superconducting-circuit platforms~\cite{ felicettiTwophotonQuantumRabi2018, felicettiUltrastrongcouplingRegimeNondipolar2018, wangStrongCouplingSinglephoton2025}. Two-photon Rabi-type models exhibit exceptional solutions and spectral collapse~\cite{duanTwophotonRabiModel2016, ricoSpectralCollapseTwophoton2020, chanBoundStatesTwophoton2020, braakSpectralDeterminantTwophoton2023, liCriticalSpectrumAnisotropic2025}, while their collective counterparts exhibit superradiant and finite-size critical behavior~\cite{garbeSuperradiantPhaseTransition2017, chenFinitesizeScalingAnalysis2018, garbeDissipationinducedBistabilityTwophoton2020}. Recent studies have shown that the superradiant regime of the two-photon DM disappears in the thermodynamic limit while spectral collapse survives~\cite{ramirezSuperradiantPhaseFinite2026}, and that the single-qubit anisotropic two-photon QRM can exhibit continuous criticality at collapse governed by a same-parity soft mode~\cite{liQuantumCriticalitySpectral2026}. These findings raise two questions: Can the one- and two-photon anisotropic Dicke models (ADMs) support exact ground-state FOTs for arbitrary finite $N$ under a common structural condition? How do their distinct bosonic responses determine the fates of these transitions---in particular, can spectral collapse terminate the two-photon FOT, and if so, does any collective signature survive in the low-energy spectrum?
	
	In this work, we construct exact Juddian ground-state doublets in the one- and two-photon ADMs, establishing symmetry-protected FOTs for arbitrary finite $N$ along their normalizable Juddian lines. A common determinant condition fixes the crossings through the product of two effective coupling strengths, whereas the individual channels set the local normal-state instability thresholds. Despite this common structure, the FOTs have distinct fates. In the one-photon ADM, coherent-branch overlap suppresses the ground-state cusp, whose strength vanishes only in the thermodynamic or infinite frequency-ratio limits, consistent with superradiant parity breaking. In the two-photon ADM, the squeezed Juddian branches lose normalizability at a finite-coupling spectral-collapse boundary, terminating the physical FOT line. As this boundary is approached along the Juddian line, the cusp strength vanishes with the collective exponent $\kappa_N=(2N-1)/4$. The same-parity soft mode retains the $N$-independent square-root scaling of collapse criticality and governs the critical response. Nevertheless, the terminating FOT leaves a collective spectral signature: the lowest different-parity gap is predicted to close with the $N$-dependent exponent $\mu_N=\kappa_N+1$. 
	
	\paragraph{Two-channel structure of exact finite-$N$ FOTs.---}
	\label{sec:model}
	
	We consider $N$ identical two-level systems coupled to a single bosonic mode,
	\begin{align}
		H_k = & \omega a^\dagger a + \Delta J_z + H_{\rm int}^{(k)}, 
		\label{eq:Hk} \\
		H_{\rm int}^{(k)} =& \frac{g}{\sqrt N} \left( a^kJ_+ + a^{\dagger k} J_- \right) + \frac{rg}{\sqrt N} \left(
		a^kJ_- + a^{\dagger k}J_+ \right). \label{eq:Hint}
	\end{align}
	Here $a$ annihilates a boson of frequency $\omega$, $\Delta$ is the two-level splitting, and $J_z$ and $J_\pm$ are collective-spin operators. We work in the fully symmetric sector $j=N/2$, with $0<r<1$. The cases $k=1, 2$ define the one- and two-photon ADMs. For a direct comparison, we adopt the $g/\sqrt N$ normalization in both models. For $k=2$, this differs from the commonly used $g/N$ convention~\cite{garbeSuperradiantPhaseTransition2017} and makes the finite-$N$ collapse threshold explicit~\cite{garbeDissipationinducedBistabilityTwophoton2020}.
	
	The interaction separates naturally into two coupling channels,
	\begin{equation}
		H_{\rm int}^{(k)} = G_+ X_k J_x + G_- Y_k J_y ,
		\label{eq:spin_channels}
	\end{equation}
	where $G_\pm=g(1\pm r)$, $X_k = (a^k + a^{\dagger k}) / \sqrt{N}$ and $Y_k = i(a^k -a^{\dagger k}) / \sqrt{N}$. The  $X_k J_x$ channel is stronger for $0<r<1$. Equivalently,
	\begin{equation}
		H_{\rm int}^{(k)} = A_k^T G(g)\mathbf J, \quad
		G(g) = g
		\begin{pmatrix}
			1&r \\ r&1
		\end{pmatrix},
		\label{eq:G}
	\end{equation}
	with $A_k = (a^k, a^{\dagger k})^T /\sqrt{N}$ and $\mathbf J = (J_+, J_-)^T$. The eigenvalues of $G$ are precisely $G_\pm$.
	
	A quadratic expansion about the fully spin-polarized normal state yields the local instability thresholds~\cite{supplement},
	\begin{equation}
		g_\pm = \frac{\sqrt{\Delta \omega}}{1 \pm r}, \quad
		g_+ < g_-.
		\label{eq:lambda_pm}
	\end{equation}
	The normal-state curvature associated with the stronger channel vanishes first at $g_+$. For the one-photon ADM, $g_+$ is the thermodynamic superradiant critical coupling~\cite{dasPhaseTransitionsAnisotropic2023}. For the two-photon ADM, $g_\pm$ instead describe local normal-state instabilities~\cite{garbeSuperradiantPhaseTransition2017, chenFinitesizeScalingAnalysis2018}, not thermodynamic phase boundaries~\cite{ramirezSuperradiantPhaseFinite2026}.
	
	The lowest exact Juddian crossing is instead fixed by the product of the channel strengths. Spin-space similarity transformations followed by displacement ($k=1$) or squeezing ($k=2$) yield the same exact Gaussian closure condition~\cite{supplement}
	\begin{equation}
		\det G(g_J) = g_J^2 (1-r^2) =\Delta \omega, \quad
		g_J^2 = \frac{\Delta \omega}{1-r^2} = g_+ g_-.
		\label{eq:det_closure}
	\end{equation}
	Hence $g_+<g_J<g_-$: the individual channels set the local instability thresholds, whereas their product locates the exact crossing. This determinant relation emerges from Gaussian boundary closure: spin-space similarity and canonical bosonic transformations triangularize the boundary-relevant couplings, thereby transferring the determinant dependence to the Juddian condition~\cite{supplement}.
	
	Both Hamiltonians conserve a generalized parity,
	\begin{equation}
		\Pi_k = \exp\left[ i\pi \left( \frac{a^\dagger a}{k} + J_z + j \right)\right], \quad
		[\Pi_k,H_k] = 0.
		\label{eq:Pik}
	\end{equation}
	For $k=2$, the lowest Juddian branch belongs to the even-boson subspace, where the parity eigenvalues are $\pm1$. At $g=g_J$, two exact Juddian eigenstates with energy $E_J^{(k)}$ are
	\begin{equation}
		\ket{\Psi_{L,R}^{(1)}} = \ket{\mp \alpha} \otimes \ket{\pm\zeta_1}_j,  \quad
		\ket{\Psi_{L,R}^{(2)}} = \ket{\mp\theta}_2 \otimes \ket{\pm\zeta_2}_j.
		\label{eq:Juddian_branches}
	\end{equation}
	Here $\ket{\pm\alpha}$ are coherent states, $\ket{\pm\theta}_2$ are squeezed vacua, and $\ket{\pm \zeta}_j \propto \exp(\pm \zeta J_-) \ket{j,j}$ are spin coherent states. The parameters are given in the Supplemental Material~\cite{supplement}. Parity exchanges the two branches, $\Pi_k \ket{\Psi_L^{(k)}} = (-1)^N \ket{\Psi_R^{(k)}}$, so the normalized parity eigenstates are
	\begin{equation}
		\ket{\Psi_{J,\pm}^{(k)}} = \frac{ \ket{\Psi_L^{(k)}} \pm (-1)^N \ket{\Psi_R^{(k)}}}{ \sqrt{2(1 \pm (-1)^N S_k)} }, \quad
		S_k = \braket{\Psi_L^{(k)}|\Psi_R^{(k)}}.
		\label{eq:Juddian_cats}
	\end{equation}
	Parity symmetry forbids hybridization of the two states at $g_J$. Their ground-state character is independently established by an exact positive-semidefinite factorization in the full spin-boson Hilbert space,
	\begin{equation}
		H_k(g_J) - E_J^{(k)} = \sum_{n=1}^{N} \mathcal Q_{n,k}^\dagger \mathcal Q_{n,k} \geqslant 0
	\end{equation} 
	with operators $\mathcal Q_{n,k}$ constructed in the Supplemental Material~\cite{supplement}. Both parity eigenstates saturate this global lower bound and are exactly degenerate ground states throughout the normalizable Juddian regime.
	
	The first-order character follows from the Juddian line
	\begin{equation}
		\Delta_J(g) = \frac{(1-r^2) g^2}{\omega},
		\label{eq:DeltaJ}
	\end{equation}
	along which the opposite-parity ground states remain exactly degenerate. Differentiating this degeneracy condition along $\Delta_J(g)$ and applying the Hellmann-Feynman theorem give the magnitude of the fixed-$\Delta$ ground-state slope discontinuity
	\begin{align}
		\mathcal J_X^{(k)} &\equiv \left| \partial_g E_{+}^{(k)} - \partial_g E_{-}^{(k)} \right|_{g=g_J}  \nonumber \\
		&= \frac{2(1-r^2)g_J}{\omega} \left|\braket{J_z}_+ - \braket{J_z}_-\right|,
		\label{eq:JX_common}
	\end{align}
	The distinct spin polarizations of the two parity sectors ensure $\mathcal J_X^{(k)} > 0$ throughout the normalizable Juddian regime~\cite{supplement}. The opposite-parity levels exchange their energy ordering at $g_J$, producing an exact finite-$N$ FOT.
	
	\paragraph{Coherent suppression versus spectral-collapse termination.---}
	\label{sec:fates}
	
\begin{figure}[t]
		\centering
		\includegraphics[width=\columnwidth]{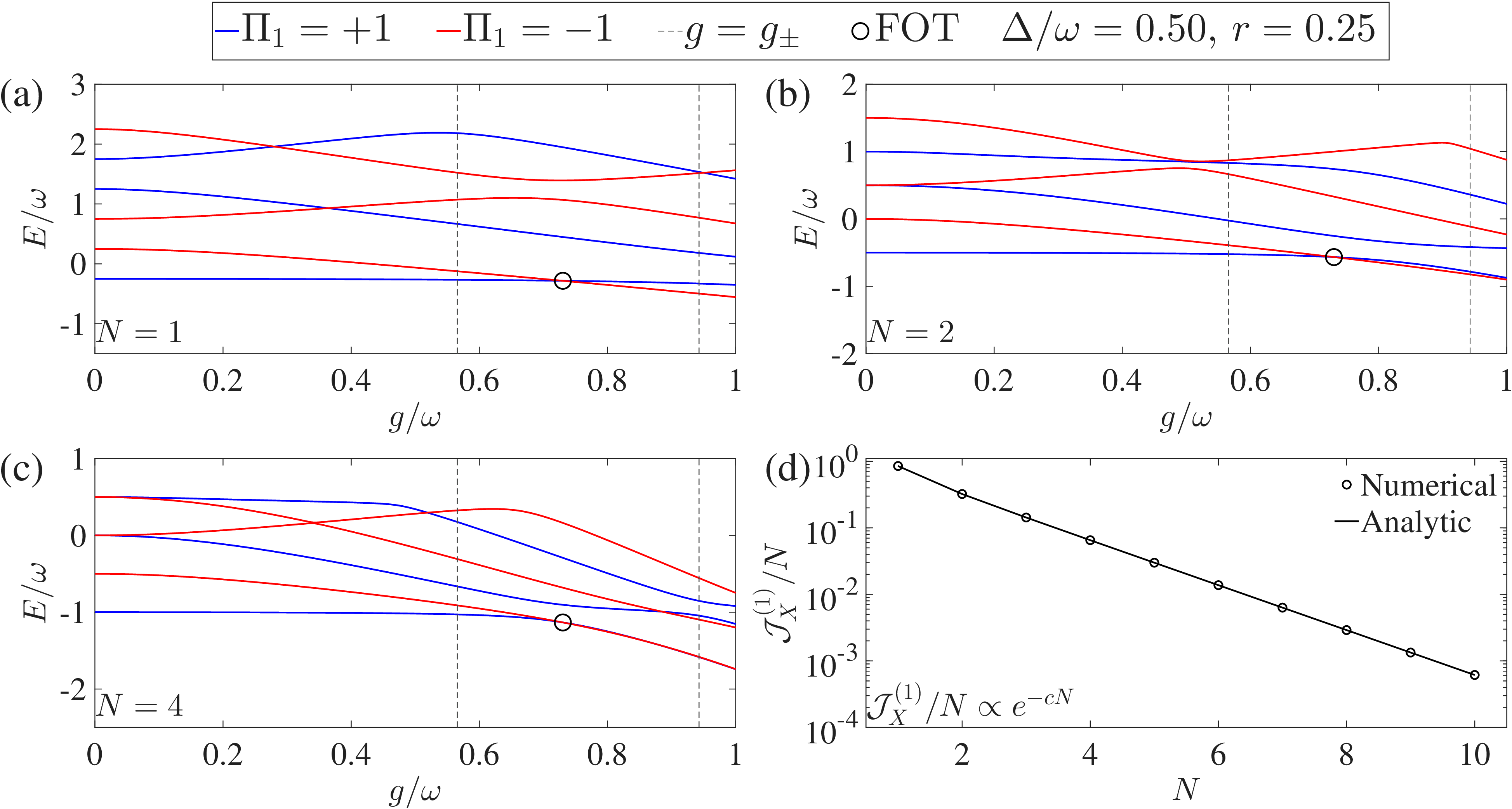}
		\caption{ 
			One-photon finite-$N$ FOT and coherent suppression.
			(a)-(c) Low-energy spectra for $r=0.25$ and $\Delta / \omega = 0.5$ at $N=1,2,4$. Open circles mark the lowest Juddian ground-state crossing at $g=g_J$, dashed lines indicate the local instability thresholds $g = g_\pm$. Blue (red) denotes positive (negative) parity.
			(d) Exponential suppression of $\mathcal J_X^{(1)}/N$:
			numerical diagonalization (circles) and the analytic result (solid line).
		}
		\label{fig:onephoton}
	\end{figure}

	Although both models share the same determinant condition for the Juddian crossing, their distinct bosonic responses lead to different fates. For the one-photon ADM, the exact states and Eq.~\eqref{eq:JX_common} give
	\begin{equation}
		\mathcal J_X^{(1)} = \frac{8Nrg_J}{\omega} \frac{|S_1|}{1-S_1^2}, \quad
		|S_1| = \left[ \frac{1-r}{1+r} e^{-\frac{2r}{1-r^2} \frac{\Delta}{\omega}} \right]^N.
		\label{eq:JX_1}
	\end{equation}
	At fixed $r$ and $\Delta/\omega>0$, the overlap of the two exact Juddian eigenstates satisfies $|S_1|=e^{-cN}$ with $c>0$. Hence
	\begin{equation}
		\mathcal J_X^{(1)} \propto Ne^{-cN} \to 0, \quad N \to \infty, \quad c>0.
		\label{eq:JX1_N}
	\end{equation}
	For every finite $N$, however, the exact Juddian ground-state doublet can be represented by parity eigenstates formed from the two coherent branches. The branch overlap vanishes as $N \to \infty$, consistent with spontaneous parity breaking in the superradiant phase~\cite{baumannExploringSymmetryBreaking2011}, and also as $\Delta/\omega \to \infty$ at fixed finite $N$, analogous to the finite-component superradiant critical limit of the QRM~\cite{hwangQuantumPhaseTransition2015}. Since $g_+<g_J$ for $0<r<1$, the Juddian crossing lies on the superradiant side of the continuous critical point. The FOT remains exact at every regular finite point on the Juddian line, but its cusp vanishes in either singular limit, leaving no separate first-order singularity within the superradiant regime. Figure~\ref{fig:onephoton} illustrates the exact ground-state crossings and the analytically derived exponential suppression of the cusp strength.

	For the two-photon ADM, by contrast, the squeezed Juddian eigenstates are normalizable only if $\sqrt N G_+(g_J)<\omega$~\cite{supplement}, restricting the physical crossing to $g_J<g_c$, where
	\begin{equation}
		g_c = \frac{\omega}{\sqrt N(1+r)}.
		\label{eq:gc}
	\end{equation}
	Projecting the Hamiltonian onto the extremal $J_x=\pm j$ spin states yields the bosonic quadratic forms
	\begin{equation}
		H_{{\rm b},\pm} = \omega a^\dagger a \pm \frac{\sqrt N}{2} G_+(g) \left(a^2 + a^{\dagger2}\right).
	\end{equation}
	The corresponding Bogoliubov frequency $\Omega_{\rm b} = \sqrt{\omega^2 - NG_+^2(g)}$ vanishes at $g_c$, signaling the loss of bosonic confinement. For $g > g_c$, the quadratic form becomes unbounded below. The stronger coupling channel therefore sets both the local instability threshold $g_+$ and the spectral-collapse threshold $g_c \propto N^{-1/2}$, independent of $\Delta$.
	
	Taking $g_J \to g_c^-$ in Eq.~\eqref{eq:det_closure} locates the endpoint at
	\begin{equation}
		\Delta_c \equiv \Delta_J(g_c) = \frac{\omega}{N}\frac{1-r}{1+r}, \quad
		\frac{g_J^2}{g_c^2} = \frac{\Delta}{\Delta_c}.
		\label{eq:Deltac}
	\end{equation}
	For $\Delta<\Delta_c$, the Juddian crossing corresponds to a normalizable ground-state doublet. At $\Delta=\Delta_c$, the squeezed states cease to be normalizable and the physical FOT line terminates. For $\Delta>\Delta_c$, the determinant condition defines only a formal continuation, without a normalizable Juddian state. The FOT window shrinks as $\Delta_c\propto N^{-1}$, consistent with the finite-size character of the two-photon superradiant regime~\cite{ramirezSuperradiantPhaseFinite2026}.
	
	\begin{figure}[t]
		\centering
		\includegraphics[width=\columnwidth]{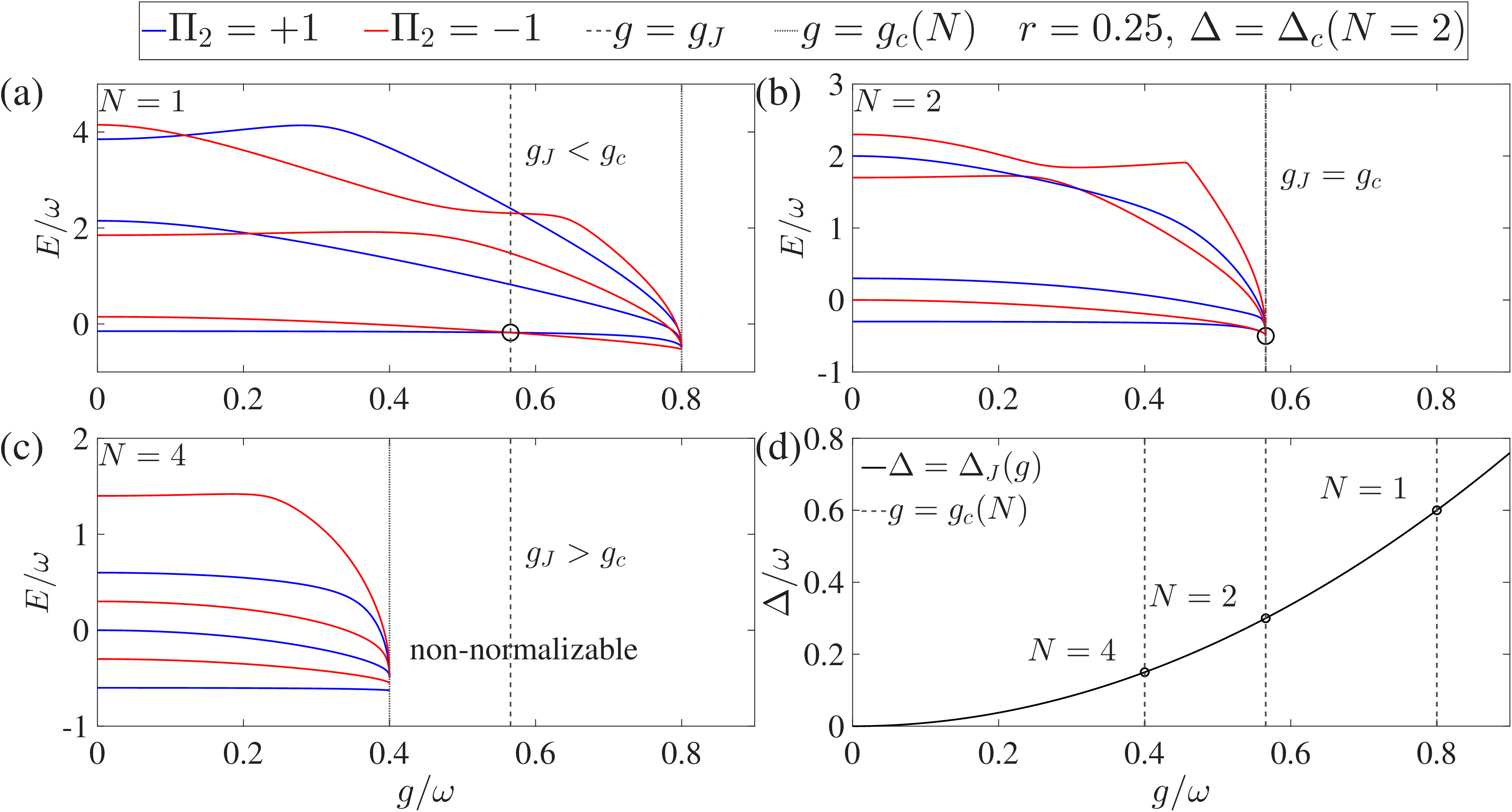}
		\caption{
			Spectral-collapse termination of the two-photon finite-$N$ FOT.
			(a)-(c) Low-energy spectra in the even-boson subspace at $r=0.25$ and $\Delta = \Delta_c(N=2) = 0.3 \omega$ for $N=1,2,4$, corresponding to $g_J<g_c$, $g_J=g_c$, and $g_J>g_c$, respectively. Blue (red) denotes positive (negative) parity. Dashed and dotted lines indicate the Juddian coupling $g_J$ and collapse threshold $g_c$, respectively. Open circles mark the physical Juddian crossing in (a) and the limiting collapse endpoint in (b). 
			(d) Common Juddian line (solid) and $N$-dependent collapse boundaries (dashed). Their intersections (open circles) locate the endpoints $(g_c, \Delta_c)$.
		}
		\label{fig:twophoton}
	\end{figure}
	
	Along the normalizable Juddian line, the two-photon cusp strength vanishes continuously as $g_J \to g_c^-$~\cite{supplement}:
	\begin{equation}
		\mathcal J_X^{(2)} \propto (g_c-g_J)^{\kappa_N}, \quad
		\kappa_N = \frac{2N-1}{4},
		\label{eq:JX_2}
	\end{equation}
	with a finite nonzero prefactor for finite $N$ and $0<r<1$~\cite{supplement}. The $N$-dependent termination exponent $\kappa_N$ originates from the collective overlap scaling of the squeezed Juddian states. In contrast, the one-photon cusp vanishes only in the thermodynamic or infinite frequency-ratio limit. As shown below, $\kappa_N$ also enters the predicted different-parity gap exponent.
	
	\paragraph{Parity splitting and soft-mode criticality at spectral collapse.---}
	\label{sec:endpoint}
	
\begin{figure}[t]
		\centering
		\includegraphics[width=\columnwidth]{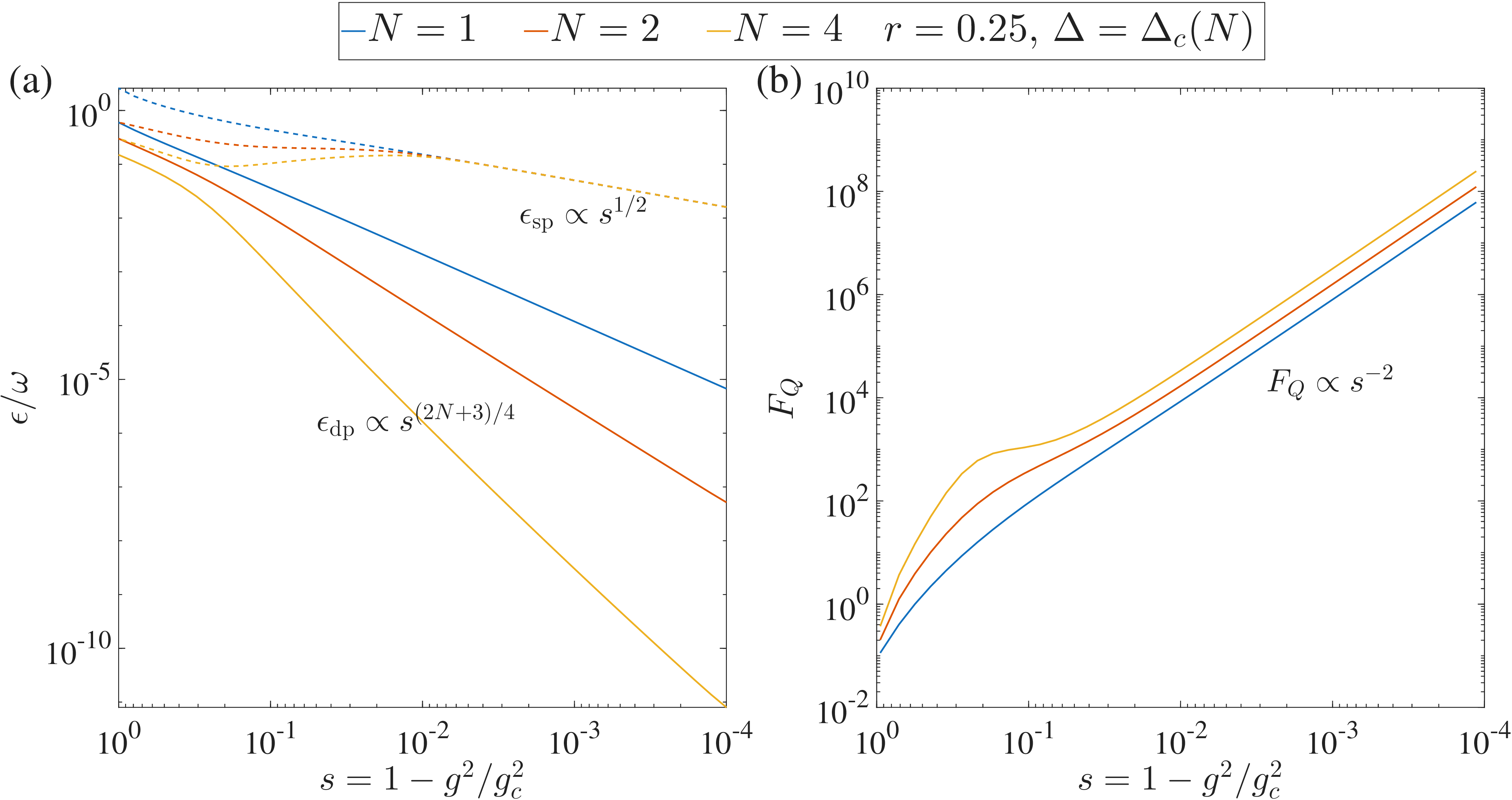}
		\caption{
			Parity-resolved endpoint gaps and critical response at $r=0.25$ and $\Delta=\Delta_c(N)$, with $s=1-g^2/g_c^2$. Blue, red, and yellow denote $N= 1, 2, 4$, respectively.
			(a) Lowest different-parity (solid) and same-parity (dashed) gaps, with exponents $\mu_N=(2N+3)/4$ and $1/2$, respectively.
			(b) Ground-state quantum Fisher information, $F_Q \propto s^{-2}$.
		}
		\label{fig:endpoint}
	\end{figure}

	Near the endpoint, detuning from the Juddian line produces a parity splitting that retains the collective $N$ dependence of the terminating FOT. At fixed $g<g_c$, a detuning $\delta \Delta \equiv \Delta-\Delta_J(g)$ introduces the parity-preserving perturbation $\delta\Delta J_z$. First-order perturbation theory gives
	\begin{equation}
		\epsilon_{\rm dp} = |\delta \Delta| \left| \braket{J_z}_+ - \braket{J_z}_- \right| + O(\delta\Delta^2),
		\label{eq:dp_gap}
	\end{equation}
	where the expectation values are taken in the exact parity-resolved Juddian states. Eqs.~\eqref{eq:JX_common} and \eqref{eq:JX_2} imply that the linear coefficient vanishes as $s^{\kappa_N}$, with $s = 1 - g^2/g_c^2 \to 0^+$. Along the endpoint trajectory $\Delta = \Delta_c$, the detuning satisfies $|\delta\Delta| = |\Delta_c - \Delta_J(g)| = \Delta_c s$. Combining this detuning with the first-order coefficient suggests the different-parity gap exponent
	\begin{equation}
		\epsilon_{\rm dp} \propto s^{\mu_N} \propto |g-g_c|^{\mu_N}, \quad 
		\mu_N = \kappa_N+1 = \frac{2N+3}{4}.
		\label{eq:muN}
	\end{equation}
	The additional power in $\mu_N=\kappa_N+1$ arises from the linear detuning away from the Juddian line, while $\kappa_N$ reflects the collective suppression of the FOT cusp. Low-energy power counting suggests that higher-order detuning corrections do not introduce a smaller exponent~\cite{supplement}. The numerical results in Fig.~\ref{fig:endpoint}(a) are consistent with the predicted gap scaling. For $N=1$, $\mu_1=5/4$ recovers the single-qubit result obtained from the diagonal adiabatic approximation~\cite{liQuantumCriticalitySpectral2026}. For $N>1$, $\mu_N$ exhibits collective $N$ dependence absent in the single-qubit result. 
	
	A distinct low-energy scale arises from a soft oscillator associated with each of the two Juddian branches. Eliminating nonextremal spin components near the endpoint gives the branch-local effective Hamiltonians~\cite{supplement}
	\begin{equation}
		H_{\rm eff}^{(L,R)} + \frac{\omega}{2} \approx \omega \left[ -\frac{4r}{(1+r)^2} \frac{\partial^2}{\partial q_{L,R}^2} + \frac{s}{4} q_{L,R}^2 \right], 
		\label{eq:Heff_endpoint}
	\end{equation}
	where $q_L = p$ and $q_R = x$. These isospectral branches are exchanged by two-photon parity. In contrast to the anisotropy-selected $x$- and $p$-type superradiant phases of the one-photon Rabi model~\cite{liuUniversalScalingCritical2017}, the two soft branches coexist at fixed $r$ and originate from the two symmetry-related Juddian branches. Their oscillator spacing satisfies $\Omega_{\rm eff}\propto s^{1/2}$. Because the Juddian doublet lies in the even-boson subspace, the lowest allowed same-parity soft excitation costs $2\Omega_{\rm eff}$, giving
	\begin{equation}
		\epsilon_{\rm sp} \propto s^{1/2} \propto |g-g_c|^{1/2}.
		\label{eq:sp_gap}
	\end{equation}
	Although the system size shifts the endpoint $(g_c,\Delta_c)$, it leaves the leading soft-mode exponent unchanged. The branch-local theory, however, does not capture the parametrically smaller interbranch parity splitting.
	
	Since $\mu_N>1/2$ for every finite $N$,
	$\epsilon_{\rm dp} \ll \epsilon_{\rm sp}$ as $g\to g_c^-$. 	As in the single-qubit collapse transition~\cite{liQuantumCriticalitySpectral2026}, the lowest excitation is therefore not the soft mode that governs the critical response. An exact parity selection rule for the ground-state quantum Fisher information (QFI) explains this distinction. Since $\Pi_2$ is independent of $g$, $[\Pi_2, \partial_g H_2] = 0$, and hence $\partial_g H_2$ cannot connect states of opposite parity. Consequently, the different-parity excitation does not contribute to the QFI spectral sum. The singular squeezing of the branch-local Gaussian ground state instead yields~\cite{supplement}
	\begin{equation}
		F_Q \propto s^{-2} \propto |g-g_c|^{-2}.
		\label{eq:FQ}
	\end{equation}
	The numerical results in Fig.~\ref{fig:endpoint}(b) agree with this $N$-independent scaling. The vanishing same-parity gap and divergent QFI reveal continuous criticality at the spectral-collapse endpoint. This singular behavior occurs even at finite $N$, as spectral collapse removes confinement in the bosonic degree of freedom.
	
	The collapse endpoint thus hosts two parametrically distinct low-energy scales: an $N$-dependent parity splitting inherited from the terminating FOT and an $N$-independent same-parity soft-mode gap governing collapse criticality. 
	
	\paragraph{Conclusion.---}
	We have established exact symmetry-protected Juddian ground-state FOTs for arbitrary finite $N$ in both one- and two-photon ADMs. In both models, the eigenvalues of the coupling matrix determine local normal-state instability thresholds, whereas its determinant fixes the exact Juddian crossing. The distinct roles of the coupling-matrix eigenvalues and determinant may also help identify exact crossings in other multichannel light-matter models. The resulting FOTs, however, have distinct fates. In the one-photon model, the overlap of displaced Juddian branches suppresses the ground-state cusp only asymptotically in the thermodynamic or infinite frequency-ratio limits, consistent with superradiant parity breaking. In the two-photon model, the squeezed Juddian branches lose normalizability at a finite-coupling spectral-collapse boundary, where the physical FOT line terminates with the collective cusp exponent $\kappa_N = (2N-1)/4$.
	
	The terminating FOT nevertheless leaves a collective imprint on the low-energy spectrum. The different-parity gap is predicted to close with the $N$-dependent exponent $\mu_N= \kappa_N+1$, linking its scaling to the vanishing FOT cusp, whereas the same-parity soft mode retains the $N$-independent square-root scaling of collapse criticality. More broadly, our results identify a route by which a symmetry-protected finite-$N$ FOT can terminate at continuous criticality while retaining its collective signature in the parity-resolved low-energy spectrum.

	\medskip
	\begin{acknowledgments}
		We acknowledge financial support from the National Key R\&D Program of China (Grants No. 2023YFA1406703 and No. 2024YFA1408900) and the National Natural Science Foundation of China under Grant No. 92565201.
	\end{acknowledgments}
	

\begin{thebibliography}{48}%
		\makeatletter
		\providecommand \@ifxundefined [1]{%
			\@ifx{#1\undefined}
		}%
		\providecommand \@ifnum [1]{%
			\ifnum #1\expandafter \@firstoftwo
			\else \expandafter \@secondoftwo
			\fi
		}%
		\providecommand \@ifx [1]{%
			\ifx #1\expandafter \@firstoftwo
			\else \expandafter \@secondoftwo
			\fi
		}%
		\providecommand \natexlab [1]{#1}%
		\providecommand \enquote  [1]{``#1''}%
		\providecommand \bibnamefont  [1]{#1}%
		\providecommand \bibfnamefont [1]{#1}%
		\providecommand \citenamefont [1]{#1}%
		\providecommand \href@noop [0]{\@secondoftwo}%
		\providecommand \href [0]{\begingroup \@sanitize@url \@href}%
		\providecommand \@href[1]{\@@startlink{#1}\@@href}%
		\providecommand \@@href[1]{\endgroup#1\@@endlink}%
		\providecommand \@sanitize@url [0]{\catcode `\\12\catcode `\$12\catcode
			`\&12\catcode `\#12\catcode `\^12\catcode `\_12\catcode `\%12\relax}%
		\providecommand \@@startlink[1]{}%
		\providecommand \@@endlink[0]{}%
		\providecommand \url  [0]{\begingroup\@sanitize@url \@url }%
		\providecommand \@url [1]{\endgroup\@href {#1}{\urlprefix }}%
		\providecommand \urlprefix  [0]{URL }%
		\providecommand \Eprint [0]{\href }%
		\providecommand \doibase [0]{https://doi.org/}%
		\providecommand \selectlanguage [0]{\@gobble}%
		\providecommand \bibinfo  [0]{\@secondoftwo}%
		\providecommand \bibfield  [0]{\@secondoftwo}%
		\providecommand \translation [1]{[#1]}%
		\providecommand \BibitemOpen [0]{}%
		\providecommand \bibitemStop [0]{}%
		\providecommand \bibitemNoStop [0]{.\EOS\space}%
		\providecommand \EOS [0]{\spacefactor3000\relax}%
		\providecommand \BibitemShut  [1]{\csname bibitem#1\endcsname}%
		\let\auto@bib@innerbib\@empty
		\bibitem [{\citenamefont {Sachdev}(2011)}]{sachdevQuantumPhaseTransitions2011}%
		\BibitemOpen
		\bibfield  {author} {\bibinfo {author} {\bibfnamefont {S.}~\bibnamefont
				{Sachdev}},\ }\href@noop {} {\emph {\bibinfo {title} {Quantum Phase
					Transitions}}},\ \bibinfo {edition} {2nd}\ ed.\ (\bibinfo  {publisher}
		{Cambridge University Press},\ \bibinfo {address} {Cambridge},\ \bibinfo
		{year} {2011})\BibitemShut {NoStop}%
		\bibitem [{\citenamefont {Vojta}(2003)}]{vojtaQuantumPhaseTransitions}%
		\BibitemOpen
		\bibfield  {author} {\bibinfo {author} {\bibfnamefont {M.}~\bibnamefont
				{Vojta}},\ }\bibfield  {title} {\bibinfo {title} {Quantum phase
				transitions},\ }\href {https://doi.org/10.1088/0034-4885/66/12/R01}
		{\bibfield  {journal} {\bibinfo  {journal} {Rep. Prog. Phys.}\ }\textbf
			{\bibinfo {volume} {66}},\ \bibinfo {pages} {2069} (\bibinfo {year}
			{2003})}\BibitemShut {NoStop}%
		\bibitem [{\citenamefont
			{Dicke}(1954)}]{dickeCoherenceSpontaneousRadiation1954a}%
		\BibitemOpen
		\bibfield  {author} {\bibinfo {author} {\bibfnamefont {R.~H.}\ \bibnamefont
				{Dicke}},\ }\bibfield  {title} {\bibinfo {title} {Coherence in spontaneous
				radiation processes},\ }\href {https://doi.org/10.1103/PhysRev.93.99}
		{\bibfield  {journal} {\bibinfo  {journal} {Phys. Rev.}\ }\textbf {\bibinfo
				{volume} {93}},\ \bibinfo {pages} {99} (\bibinfo {year} {1954})}\BibitemShut
		{NoStop}%
		\bibitem [{\citenamefont {Hepp}\ and\ \citenamefont
			{Lieb}(1973)}]{heppEquilibriumStatisticalMechanics1973}%
		\BibitemOpen
		\bibfield  {author} {\bibinfo {author} {\bibfnamefont {K.}~\bibnamefont
				{Hepp}}\ and\ \bibinfo {author} {\bibfnamefont {E.~H.}\ \bibnamefont
				{Lieb}},\ }\bibfield  {title} {\bibinfo {title} {Equilibrium statistical
				mechanics of matter interacting with the quantized radiation field},\ }\href
		{https://doi.org/10.1103/PhysRevA.8.2517} {\bibfield  {journal} {\bibinfo
				{journal} {Phys. Rev. A}\ }\textbf {\bibinfo {volume} {8}},\ \bibinfo {pages}
			{2517} (\bibinfo {year} {1973})}\BibitemShut {NoStop}%
		\bibitem [{\citenamefont {Wang}\ and\ \citenamefont
			{Hioe}(1973)}]{wangPhaseTransitionDicke1973}%
		\BibitemOpen
		\bibfield  {author} {\bibinfo {author} {\bibfnamefont {Y.~K.}\ \bibnamefont
				{Wang}}\ and\ \bibinfo {author} {\bibfnamefont {F.~T.}\ \bibnamefont
				{Hioe}},\ }\bibfield  {title} {\bibinfo {title} {Phase transition in the
				{Dicke} model of superradiance},\ }\href
		{https://doi.org/10.1103/PhysRevA.7.831} {\bibfield  {journal} {\bibinfo
				{journal} {Phys. Rev. A}\ }\textbf {\bibinfo {volume} {7}},\ \bibinfo {pages}
			{831} (\bibinfo {year} {1973})}\BibitemShut {NoStop}%
		\bibitem [{\citenamefont {Emary}\ and\ \citenamefont
			{Brandes}(2003)}]{emaryQuantumChaosTriggered2003}%
		\BibitemOpen
		\bibfield  {author} {\bibinfo {author} {\bibfnamefont {C.}~\bibnamefont
				{Emary}}\ and\ \bibinfo {author} {\bibfnamefont {T.}~\bibnamefont
				{Brandes}},\ }\bibfield  {title} {\bibinfo {title} {Quantum chaos triggered
				by precursors of a quantum phase transition: The {Dicke} model},\ }\href
		{https://doi.org/10.1103/PhysRevLett.90.044101} {\bibfield  {journal}
			{\bibinfo  {journal} {Phys. Rev. Lett.}\ }\textbf {\bibinfo {volume} {90}},\
			\bibinfo {pages} {044101} (\bibinfo {year} {2003})}\BibitemShut {NoStop}%
		\bibitem [{\citenamefont {Kirton}\ \emph {et~al.}(2019)\citenamefont {Kirton},
			\citenamefont {Roses}, \citenamefont {Keeling},\ and\ \citenamefont
			{Dalla~Torre}}]{kirtonIntroductionDickeModel2019}%
		\BibitemOpen
		\bibfield  {author} {\bibinfo {author} {\bibfnamefont {P.}~\bibnamefont
				{Kirton}}, \bibinfo {author} {\bibfnamefont {M.~M.}\ \bibnamefont {Roses}},
			\bibinfo {author} {\bibfnamefont {J.}~\bibnamefont {Keeling}},\ and\ \bibinfo
			{author} {\bibfnamefont {E.~G.}\ \bibnamefont {Dalla~Torre}},\ }\bibfield
		{title} {\bibinfo {title} {Introduction to the {Dicke} model: From
				equilibrium to nonequilibrium, and \emph{vice versa}},\ }\href
		{https://doi.org/10.1002/qute.201800043} {\bibfield  {journal} {\bibinfo
				{journal} {Adv. Quantum Technol.}\ }\textbf {\bibinfo {volume} {2}},\
			\bibinfo {pages} {1800043} (\bibinfo {year} {2019})}\BibitemShut {NoStop}%
		\bibitem [{\citenamefont {Baumann}\ \emph {et~al.}(2010)\citenamefont
			{Baumann}, \citenamefont {Guerlin}, \citenamefont {Brennecke},\ and\
			\citenamefont {Esslinger}}]{baumannDickeQuantumPhase2010}%
		\BibitemOpen
		\bibfield  {author} {\bibinfo {author} {\bibfnamefont {K.}~\bibnamefont
				{Baumann}}, \bibinfo {author} {\bibfnamefont {C.}~\bibnamefont {Guerlin}},
			\bibinfo {author} {\bibfnamefont {F.}~\bibnamefont {Brennecke}},\ and\
			\bibinfo {author} {\bibfnamefont {T.}~\bibnamefont {Esslinger}},\ }\bibfield
		{title} {\bibinfo {title} {{Dicke} quantum phase transition with a superfluid
				gas in an optical cavity},\ }\href {https://doi.org/10.1038/nature09009}
		{\bibfield  {journal} {\bibinfo  {journal} {Nature}\ }\textbf {\bibinfo
				{volume} {464}},\ \bibinfo {pages} {1301} (\bibinfo {year}
			{2010})}\BibitemShut {NoStop}%
		\bibitem [{\citenamefont {Nagy}\ \emph {et~al.}(2010)\citenamefont {Nagy},
			\citenamefont {K{\'o}nya}, \citenamefont {Szirmai},\ and\ \citenamefont
			{Domokos}}]{nagyDickemodelPhaseTransition2010}%
		\BibitemOpen
		\bibfield  {author} {\bibinfo {author} {\bibfnamefont {D.}~\bibnamefont
				{Nagy}}, \bibinfo {author} {\bibfnamefont {G.}~\bibnamefont {K{\'o}nya}},
			\bibinfo {author} {\bibfnamefont {G.}~\bibnamefont {Szirmai}},\ and\ \bibinfo
			{author} {\bibfnamefont {P.}~\bibnamefont {Domokos}},\ }\bibfield  {title}
		{\bibinfo {title} {{Dicke}-model phase transition in the quantum motion of a
				{Bose-Einstein} condensate in an optical cavity},\ }\href
		{https://doi.org/10.1103/PhysRevLett.104.130401} {\bibfield  {journal}
			{\bibinfo  {journal} {Phys. Rev. Lett.}\ }\textbf {\bibinfo {volume} {104}},\
			\bibinfo {pages} {130401} (\bibinfo {year} {2010})}\BibitemShut {NoStop}%
		\bibitem [{\citenamefont {Zhang}\ \emph {et~al.}(2018)\citenamefont {Zhang},
			\citenamefont {Lee}, \citenamefont {Kumar}, \citenamefont {Arnold},
			\citenamefont {Masson}, \citenamefont {Grimsmo}, \citenamefont {Parkins},\
			and\ \citenamefont {Barrett}}]{zhangDickemodelSimulationCavityassisted2018}%
		\BibitemOpen
		\bibfield  {author} {\bibinfo {author} {\bibfnamefont {Z.}~\bibnamefont
				{Zhang}}, \bibinfo {author} {\bibfnamefont {C.~H.}\ \bibnamefont {Lee}},
			\bibinfo {author} {\bibfnamefont {R.}~\bibnamefont {Kumar}}, \bibinfo
			{author} {\bibfnamefont {K.~J.}\ \bibnamefont {Arnold}}, \bibinfo {author}
			{\bibfnamefont {S.~J.}\ \bibnamefont {Masson}}, \bibinfo {author}
			{\bibfnamefont {A.~L.}\ \bibnamefont {Grimsmo}}, \bibinfo {author}
			{\bibfnamefont {A.~S.}\ \bibnamefont {Parkins}},\ and\ \bibinfo {author}
			{\bibfnamefont {M.~D.}\ \bibnamefont {Barrett}},\ }\bibfield  {title}
		{\bibinfo {title} {{Dicke}-model simulation via cavity-assisted {Raman}
				transitions},\ }\href {https://doi.org/10.1103/PhysRevA.97.043858} {\bibfield
			{journal} {\bibinfo  {journal} {Phys. Rev. A}\ }\textbf {\bibinfo {volume}
				{97}},\ \bibinfo {pages} {043858} (\bibinfo {year} {2018})}\BibitemShut
		{NoStop}%
		\bibitem [{\citenamefont
			{Ashhab}(2013)}]{ashhabSuperradianceTransitionSystem2013}%
		\BibitemOpen
		\bibfield  {author} {\bibinfo {author} {\bibfnamefont {S.}~\bibnamefont
				{Ashhab}},\ }\bibfield  {title} {\bibinfo {title} {Superradiance transition
				in a system with a single qubit and a single oscillator},\ }\href
		{https://doi.org/10.1103/PhysRevA.87.013826} {\bibfield  {journal} {\bibinfo
				{journal} {Phys. Rev. A}\ }\textbf {\bibinfo {volume} {87}},\ \bibinfo
			{pages} {013826} (\bibinfo {year} {2013})}\BibitemShut {NoStop}%
		\bibitem [{\citenamefont {Hwang}\ \emph {et~al.}(2015)\citenamefont {Hwang},
			\citenamefont {Puebla},\ and\ \citenamefont
			{Plenio}}]{hwangQuantumPhaseTransition2015}%
		\BibitemOpen
		\bibfield  {author} {\bibinfo {author} {\bibfnamefont {M.-J.}\ \bibnamefont
				{Hwang}}, \bibinfo {author} {\bibfnamefont {R.}~\bibnamefont {Puebla}},\ and\
			\bibinfo {author} {\bibfnamefont {M.~B.}\ \bibnamefont {Plenio}},\ }\bibfield
		{title} {\bibinfo {title} {Quantum phase transition and universal dynamics
				in the {Rabi} model},\ }\href
		{https://doi.org/10.1103/PhysRevLett.115.180404} {\bibfield  {journal}
			{\bibinfo  {journal} {Phys. Rev. Lett.}\ }\textbf {\bibinfo {volume} {115}},\
			\bibinfo {pages} {180404} (\bibinfo {year} {2015})}\BibitemShut {NoStop}%
		\bibitem [{\citenamefont {Felicetti}\ and\ \citenamefont
			{Le~Boit{\'e}}(2020)}]{felicettiUniversalSpectralFeatures2020}%
		\BibitemOpen
		\bibfield  {author} {\bibinfo {author} {\bibfnamefont {S.}~\bibnamefont
				{Felicetti}}\ and\ \bibinfo {author} {\bibfnamefont {A.}~\bibnamefont
				{Le~Boit{\'e}}},\ }\bibfield  {title} {\bibinfo {title} {Universal spectral
				features of ultrastrongly coupled systems},\ }\href
		{https://doi.org/10.1103/PhysRevLett.124.040404} {\bibfield  {journal}
			{\bibinfo  {journal} {Phys. Rev. Lett.}\ }\textbf {\bibinfo {volume} {124}},\
			\bibinfo {pages} {040404} (\bibinfo {year} {2020})}\BibitemShut {NoStop}%
		\bibitem [{\citenamefont {Garbe}\ \emph
			{et~al.}(2020{\natexlab{a}})\citenamefont {Garbe}, \citenamefont {Bina},
			\citenamefont {Keller}, \citenamefont {Paris},\ and\ \citenamefont
			{Felicetti}}]{garbeCriticalQuantumMetrology2020a}%
		\BibitemOpen
		\bibfield  {author} {\bibinfo {author} {\bibfnamefont {L.}~\bibnamefont
				{Garbe}}, \bibinfo {author} {\bibfnamefont {M.}~\bibnamefont {Bina}},
			\bibinfo {author} {\bibfnamefont {A.}~\bibnamefont {Keller}}, \bibinfo
			{author} {\bibfnamefont {M.~G.~A.}\ \bibnamefont {Paris}},\ and\ \bibinfo
			{author} {\bibfnamefont {S.}~\bibnamefont {Felicetti}},\ }\bibfield  {title}
		{\bibinfo {title} {Critical quantum metrology with a finite-component quantum
				phase transition},\ }\href {https://doi.org/10.1103/PhysRevLett.124.120504}
		{\bibfield  {journal} {\bibinfo  {journal} {Phys. Rev. Lett.}\ }\textbf
			{\bibinfo {volume} {124}},\ \bibinfo {pages} {120504} (\bibinfo {year}
			{2020}{\natexlab{a}})}\BibitemShut {NoStop}%
		\bibitem [{\citenamefont {Cai}\ \emph {et~al.}(2021)\citenamefont {Cai},
			\citenamefont {Liu}, \citenamefont {Zhao}, \citenamefont {Wu}, \citenamefont
			{Mei}, \citenamefont {Jiang}, \citenamefont {He}, \citenamefont {Zhang},
			\citenamefont {Zhou},\ and\ \citenamefont
			{Duan}}]{caiObservationQuantumPhase2021}%
		\BibitemOpen
		\bibfield  {author} {\bibinfo {author} {\bibfnamefont {M.-L.}\ \bibnamefont
				{Cai}}, \bibinfo {author} {\bibfnamefont {Z.-D.}\ \bibnamefont {Liu}},
			\bibinfo {author} {\bibfnamefont {W.-D.}\ \bibnamefont {Zhao}}, \bibinfo
			{author} {\bibfnamefont {Y.-K.}\ \bibnamefont {Wu}}, \bibinfo {author}
			{\bibfnamefont {Q.-X.}\ \bibnamefont {Mei}}, \bibinfo {author} {\bibfnamefont
				{Y.}~\bibnamefont {Jiang}}, \bibinfo {author} {\bibfnamefont
				{L.}~\bibnamefont {He}}, \bibinfo {author} {\bibfnamefont {X.}~\bibnamefont
				{Zhang}}, \bibinfo {author} {\bibfnamefont {Z.-C.}\ \bibnamefont {Zhou}},\
			and\ \bibinfo {author} {\bibfnamefont {L.-M.}\ \bibnamefont {Duan}},\
		}\bibfield  {title} {\bibinfo {title} {Observation of a quantum phase
				transition in the quantum {Rabi} model with a single trapped ion},\ }\href
		{https://doi.org/10.1038/s41467-021-21425-8} {\bibfield  {journal} {\bibinfo
				{journal} {Nat. Commun.}\ }\textbf {\bibinfo {volume} {12}},\ \bibinfo
			{pages} {1126} (\bibinfo {year} {2021})}\BibitemShut {NoStop}%
		\bibitem [{\citenamefont {Chen}\ \emph
			{et~al.}(2021{\natexlab{a}})\citenamefont {Chen}, \citenamefont {Wu},
			\citenamefont {Jiang}, \citenamefont {L{\"u}}, \citenamefont {Peng},\ and\
			\citenamefont {Du}}]{chenExperimentalQuantumSimulation2021a}%
		\BibitemOpen
		\bibfield  {author} {\bibinfo {author} {\bibfnamefont {X.}~\bibnamefont
				{Chen}}, \bibinfo {author} {\bibfnamefont {Z.}~\bibnamefont {Wu}}, \bibinfo
			{author} {\bibfnamefont {M.}~\bibnamefont {Jiang}}, \bibinfo {author}
			{\bibfnamefont {X.-Y.}\ \bibnamefont {L{\"u}}}, \bibinfo {author}
			{\bibfnamefont {X.}~\bibnamefont {Peng}},\ and\ \bibinfo {author}
			{\bibfnamefont {J.}~\bibnamefont {Du}},\ }\bibfield  {title} {\bibinfo
			{title} {Experimental quantum simulation of superradiant phase transition
				beyond no-go theorem via antisqueezing},\ }\href
		{https://doi.org/10.1038/s41467-021-26573-5} {\bibfield  {journal} {\bibinfo
				{journal} {Nat. Commun.}\ }\textbf {\bibinfo {volume} {12}},\ \bibinfo
			{pages} {6281} (\bibinfo {year} {2021}{\natexlab{a}})}\BibitemShut {NoStop}%
		\bibitem [{\citenamefont {Arias}\ \emph {et~al.}(2003)\citenamefont {Arias},
			\citenamefont {Dukelsky},\ and\ \citenamefont
			{{Garc{\'i}a-Ramos}}}]{ariasQuantumPhaseTransitions2003}%
		\BibitemOpen
		\bibfield  {author} {\bibinfo {author} {\bibfnamefont {J.~M.}\ \bibnamefont
				{Arias}}, \bibinfo {author} {\bibfnamefont {J.}~\bibnamefont {Dukelsky}},\
			and\ \bibinfo {author} {\bibfnamefont {J.~E.}\ \bibnamefont
				{{Garc{\'i}a-Ramos}}},\ }\bibfield  {title} {\bibinfo {title} {Quantum phase
				transitions in the interacting boson model: Integrability, level repulsion,
				and level crossing},\ }\href {https://doi.org/10.1103/PhysRevLett.91.162502}
		{\bibfield  {journal} {\bibinfo  {journal} {Phys. Rev. Lett.}\ }\textbf
			{\bibinfo {volume} {91}},\ \bibinfo {pages} {162502} (\bibinfo {year}
			{2003})}\BibitemShut {NoStop}%
		\bibitem [{\citenamefont {Wu}\ \emph {et~al.}(2004)\citenamefont {Wu},
			\citenamefont {Sarandy},\ and\ \citenamefont
			{Lidar}}]{wuQuantumPhaseTransitions2004}%
		\BibitemOpen
		\bibfield  {author} {\bibinfo {author} {\bibfnamefont {L.-A.}\ \bibnamefont
				{Wu}}, \bibinfo {author} {\bibfnamefont {M.~S.}\ \bibnamefont {Sarandy}},\
			and\ \bibinfo {author} {\bibfnamefont {D.~A.}\ \bibnamefont {Lidar}},\
		}\bibfield  {title} {\bibinfo {title} {Quantum phase transitions and
				bipartite entanglement},\ }\href
		{https://doi.org/10.1103/PhysRevLett.93.250404} {\bibfield  {journal}
			{\bibinfo  {journal} {Phys. Rev. Lett.}\ }\textbf {\bibinfo {volume} {93}},\
			\bibinfo {pages} {250404} (\bibinfo {year} {2004})}\BibitemShut {NoStop}%
		\bibitem [{\citenamefont {Judd}(1979)}]{juddExactSolutionsClass1979a}%
		\BibitemOpen
		\bibfield  {author} {\bibinfo {author} {\bibfnamefont {B.~R.}\ \bibnamefont
				{Judd}},\ }\bibfield  {title} {\bibinfo {title} {Exact solutions to a class
				of {{Jahn-Teller}} systems},\ }\href
		{https://doi.org/10.1088/0022-3719/12/9/010} {\bibfield  {journal} {\bibinfo
				{journal} {J. Phys. C: Solid State Phys.}\ }\textbf {\bibinfo {volume}
				{12}},\ \bibinfo {pages} {1685} (\bibinfo {year} {1979})}\BibitemShut
		{NoStop}%
		\bibitem [{\citenamefont {Emary}\ and\ \citenamefont
			{Bishop}(2002)}]{emaryExactIsolatedSolutions2002}%
		\BibitemOpen
		\bibfield  {author} {\bibinfo {author} {\bibfnamefont {C.}~\bibnamefont
				{Emary}}\ and\ \bibinfo {author} {\bibfnamefont {R.~F.}\ \bibnamefont
				{Bishop}},\ }\bibfield  {title} {\bibinfo {title} {Exact isolated solutions
				for the two-photon {Rabi} {Hamiltonian}},\ }\href
		{https://doi.org/10.1088/0305-4470/35/39/307} {\bibfield  {journal} {\bibinfo
				{journal} {J. Phys. A: Math. Gen.}\ }\textbf {\bibinfo {volume} {35}},\
			\bibinfo {pages} {8231} (\bibinfo {year} {2002})}\BibitemShut {NoStop}%
		\bibitem [{\citenamefont {Braak}(2011)}]{braakIntegrabilityRabiModel2011}%
		\BibitemOpen
		\bibfield  {author} {\bibinfo {author} {\bibfnamefont {D.}~\bibnamefont
				{Braak}},\ }\bibfield  {title} {\bibinfo {title} {Integrability of the {Rabi}
				model},\ }\href {https://doi.org/10.1103/PhysRevLett.107.100401} {\bibfield
			{journal} {\bibinfo  {journal} {Phys. Rev. Lett.}\ }\textbf {\bibinfo
				{volume} {107}},\ \bibinfo {pages} {100401} (\bibinfo {year}
			{2011})}\BibitemShut {NoStop}%
		\bibitem [{\citenamefont {Chen}\ \emph {et~al.}(2012)\citenamefont {Chen},
			\citenamefont {Wang}, \citenamefont {He}, \citenamefont {Liu},\ and\
			\citenamefont {Wang}}]{chenExactSolvabilityQuantum2012}%
		\BibitemOpen
		\bibfield  {author} {\bibinfo {author} {\bibfnamefont {Q.-H.}\ \bibnamefont
				{Chen}}, \bibinfo {author} {\bibfnamefont {C.}~\bibnamefont {Wang}}, \bibinfo
			{author} {\bibfnamefont {S.}~\bibnamefont {He}}, \bibinfo {author}
			{\bibfnamefont {T.}~\bibnamefont {Liu}},\ and\ \bibinfo {author}
			{\bibfnamefont {K.-L.}\ \bibnamefont {Wang}},\ }\bibfield  {title} {\bibinfo
			{title} {Exact solvability of the quantum {Rabi} model using {Bogoliubov}
				operators},\ }\href {https://doi.org/10.1103/PhysRevA.86.023822} {\bibfield
			{journal} {\bibinfo  {journal} {Phys. Rev. A}\ }\textbf {\bibinfo {volume}
				{86}},\ \bibinfo {pages} {023822} (\bibinfo {year} {2012})}\BibitemShut
		{NoStop}%
		\bibitem [{\citenamefont {Xie}\ \emph {et~al.}(2014)\citenamefont {Xie},
			\citenamefont {Cui}, \citenamefont {Cao}, \citenamefont {Amico},\ and\
			\citenamefont {Fan}}]{xieAnisotropicRabiModel2014}%
		\BibitemOpen
		\bibfield  {author} {\bibinfo {author} {\bibfnamefont {Q.-T.}\ \bibnamefont
				{Xie}}, \bibinfo {author} {\bibfnamefont {S.}~\bibnamefont {Cui}}, \bibinfo
			{author} {\bibfnamefont {J.-P.}\ \bibnamefont {Cao}}, \bibinfo {author}
			{\bibfnamefont {L.}~\bibnamefont {Amico}},\ and\ \bibinfo {author}
			{\bibfnamefont {H.}~\bibnamefont {Fan}},\ }\bibfield  {title} {\bibinfo
			{title} {Anisotropic {Rabi} model},\ }\href
		{https://doi.org/10.1103/PhysRevX.4.021046} {\bibfield  {journal} {\bibinfo
				{journal} {Phys. Rev. X}\ }\textbf {\bibinfo {volume} {4}},\ \bibinfo {pages}
			{021046} (\bibinfo {year} {2014})}\BibitemShut {NoStop}%
		\bibitem [{\citenamefont {Chen}\ \emph
			{et~al.}(2021{\natexlab{b}})\citenamefont {Chen}, \citenamefont {Duan},
			\citenamefont {Braak},\ and\ \citenamefont
			{Chen}}]{chenMultipleGroundstateInstabilities2021}%
		\BibitemOpen
		\bibfield  {author} {\bibinfo {author} {\bibfnamefont {X.-Y.}\ \bibnamefont
				{Chen}}, \bibinfo {author} {\bibfnamefont {L.}~\bibnamefont {Duan}}, \bibinfo
			{author} {\bibfnamefont {D.}~\bibnamefont {Braak}},\ and\ \bibinfo {author}
			{\bibfnamefont {Q.-H.}\ \bibnamefont {Chen}},\ }\bibfield  {title} {\bibinfo
			{title} {Multiple ground-state instabilities in the anisotropic quantum
				{Rabi} model},\ }\href {https://doi.org/10.1103/PhysRevA.103.043708}
		{\bibfield  {journal} {\bibinfo  {journal} {Phys. Rev. A}\ }\textbf {\bibinfo
				{volume} {103}},\ \bibinfo {pages} {043708} (\bibinfo {year}
			{2021}{\natexlab{b}})}\BibitemShut {NoStop}%
		\bibitem [{\citenamefont {He}\ \emph {et~al.}(2015)\citenamefont {He},
			\citenamefont {Duan},\ and\ \citenamefont
			{Chen}}]{heExactSolvabilityNonintegrability2015}%
		\BibitemOpen
		\bibfield  {author} {\bibinfo {author} {\bibfnamefont {S.}~\bibnamefont
				{He}}, \bibinfo {author} {\bibfnamefont {L.}~\bibnamefont {Duan}},\ and\
			\bibinfo {author} {\bibfnamefont {Q.-H.}\ \bibnamefont {Chen}},\ }\bibfield
		{title} {\bibinfo {title} {Exact solvability, non-integrability, and genuine
				multipartite entanglement dynamics of the {{Dicke}} model},\ }\href
		{https://doi.org/10.1088/1367-2630/17/4/043033} {\bibfield  {journal}
			{\bibinfo  {journal} {New J. Phys.}\ }\textbf {\bibinfo {volume} {17}},\
			\bibinfo {pages} {043033} (\bibinfo {year} {2015})}\BibitemShut {NoStop}%
		\bibitem [{\citenamefont {Buijsman}\ \emph {et~al.}(2017)\citenamefont
			{Buijsman}, \citenamefont {Gritsev},\ and\ \citenamefont
			{Sprik}}]{buijsmanNonergodicityAnisotropicDicke2017}%
		\BibitemOpen
		\bibfield  {author} {\bibinfo {author} {\bibfnamefont {W.}~\bibnamefont
				{Buijsman}}, \bibinfo {author} {\bibfnamefont {V.}~\bibnamefont {Gritsev}},\
			and\ \bibinfo {author} {\bibfnamefont {R.}~\bibnamefont {Sprik}},\ }\bibfield
		{title} {\bibinfo {title} {Nonergodicity in the anisotropic {Dicke} model},\
		}\href {https://doi.org/10.1103/PhysRevLett.118.080601} {\bibfield  {journal}
			{\bibinfo  {journal} {Phys. Rev. Lett.}\ }\textbf {\bibinfo {volume} {118}},\
			\bibinfo {pages} {080601} (\bibinfo {year} {2017})}\BibitemShut {NoStop}%
		\bibitem [{\citenamefont {Liu}\ \emph {et~al.}(2017)\citenamefont {Liu},
			\citenamefont {Chesi}, \citenamefont {Ying}, \citenamefont {Chen},
			\citenamefont {Luo},\ and\ \citenamefont
			{Lin}}]{liuUniversalScalingCritical2017}%
		\BibitemOpen
		\bibfield  {author} {\bibinfo {author} {\bibfnamefont {M.}~\bibnamefont
				{Liu}}, \bibinfo {author} {\bibfnamefont {S.}~\bibnamefont {Chesi}}, \bibinfo
			{author} {\bibfnamefont {Z.-J.}\ \bibnamefont {Ying}}, \bibinfo {author}
			{\bibfnamefont {X.}~\bibnamefont {Chen}}, \bibinfo {author} {\bibfnamefont
				{H.-G.}\ \bibnamefont {Luo}},\ and\ \bibinfo {author} {\bibfnamefont {H.-Q.}\
				\bibnamefont {Lin}},\ }\bibfield  {title} {\bibinfo {title} {Universal
				scaling and critical exponents of the anisotropic quantum {Rabi} model},\
		}\href {https://doi.org/10.1103/PhysRevLett.119.220601} {\bibfield  {journal}
			{\bibinfo  {journal} {Phys. Rev. Lett.}\ }\textbf {\bibinfo {volume} {119}},\
			\bibinfo {pages} {220601} (\bibinfo {year} {2017})}\BibitemShut {NoStop}%
		\bibitem [{\citenamefont {Li}\ \emph {et~al.}(2022)\citenamefont {Li},
			\citenamefont {Fazio},\ and\ \citenamefont
			{Chesi}}]{liNonlinearDynamicsDissipative2022}%
		\BibitemOpen
		\bibfield  {author} {\bibinfo {author} {\bibfnamefont {J.}~\bibnamefont
				{Li}}, \bibinfo {author} {\bibfnamefont {R.}~\bibnamefont {Fazio}},\ and\
			\bibinfo {author} {\bibfnamefont {S.}~\bibnamefont {Chesi}},\ }\bibfield
		{title} {\bibinfo {title} {Nonlinear dynamics of the dissipative anisotropic
				two-photon {Dicke} model},\ }\href {https://doi.org/10.1088/1367-2630/ac8897}
		{\bibfield  {journal} {\bibinfo  {journal} {New J. Phys.}\ }\textbf {\bibinfo
				{volume} {24}},\ \bibinfo {pages} {083039} (\bibinfo {year}
			{2022})}\BibitemShut {NoStop}%
		\bibitem [{\citenamefont {Das}\ \emph {et~al.}(2023)\citenamefont {Das},
			\citenamefont {Bhakuni},\ and\ \citenamefont
			{Sharma}}]{dasPhaseTransitionsAnisotropic2023}%
		\BibitemOpen
		\bibfield  {author} {\bibinfo {author} {\bibfnamefont {P.}~\bibnamefont
				{Das}}, \bibinfo {author} {\bibfnamefont {D.~S.}\ \bibnamefont {Bhakuni}},\
			and\ \bibinfo {author} {\bibfnamefont {A.}~\bibnamefont {Sharma}},\
		}\bibfield  {title} {\bibinfo {title} {Phase transitions of the anisotropic
				{Dicke} model},\ }\href {https://doi.org/10.1103/PhysRevA.107.043706}
		{\bibfield  {journal} {\bibinfo  {journal} {Phys. Rev. A}\ }\textbf {\bibinfo
				{volume} {107}},\ \bibinfo {pages} {043706} (\bibinfo {year}
			{2023})}\BibitemShut {NoStop}%
		\bibitem [{\citenamefont {He}\ and\ \citenamefont
			{Song}(2026)}]{heHierarchicSuperradiantPhases2026a}%
		\BibitemOpen
		\bibfield  {author} {\bibinfo {author} {\bibfnamefont {D.}~\bibnamefont
				{He}}\ and\ \bibinfo {author} {\bibfnamefont {Z.}~\bibnamefont {Song}},\
		}\bibfield  {title} {\bibinfo {title} {Hierarchic superradiant phases in
				anisotropic {Dicke} model},\ }\href
		{https://doi.org/10.21468/SciPostPhysCore.9.1.009} {\bibfield  {journal}
			{\bibinfo  {journal} {SciPost Phys. Core}\ }\textbf {\bibinfo {volume} {9}},\
			\bibinfo {pages} {9} (\bibinfo {year} {2026})}\BibitemShut {NoStop}%
		\bibitem [{\citenamefont {Felicetti}\ \emph {et~al.}(2015)\citenamefont
			{Felicetti}, \citenamefont {Pedernales}, \citenamefont {Egusquiza},
			\citenamefont {Romero}, \citenamefont {Lamata}, \citenamefont {Braak},\ and\
			\citenamefont {Solano}}]{felicettiSpectralCollapseTwophonon2015}%
		\BibitemOpen
		\bibfield  {author} {\bibinfo {author} {\bibfnamefont {S.}~\bibnamefont
				{Felicetti}}, \bibinfo {author} {\bibfnamefont {J.~S.}\ \bibnamefont
				{Pedernales}}, \bibinfo {author} {\bibfnamefont {I.~L.}\ \bibnamefont
				{Egusquiza}}, \bibinfo {author} {\bibfnamefont {G.}~\bibnamefont {Romero}},
			\bibinfo {author} {\bibfnamefont {L.}~\bibnamefont {Lamata}}, \bibinfo
			{author} {\bibfnamefont {D.}~\bibnamefont {Braak}},\ and\ \bibinfo {author}
			{\bibfnamefont {E.}~\bibnamefont {Solano}},\ }\bibfield  {title} {\bibinfo
			{title} {Spectral collapse via two-phonon interactions in trapped ions},\
		}\href {https://doi.org/10.1103/PhysRevA.92.033817} {\bibfield  {journal}
			{\bibinfo  {journal} {Phys. Rev. A}\ }\textbf {\bibinfo {volume} {92}},\
			\bibinfo {pages} {033817} (\bibinfo {year} {2015})}\BibitemShut {NoStop}%
		\bibitem [{\citenamefont {Casanova}\ \emph {et~al.}(2018)\citenamefont
			{Casanova}, \citenamefont {Puebla}, \citenamefont {{Moya-Cessa}},\ and\
			\citenamefont {Plenio}}]{casanovaConnectingNthOrder2018}%
		\BibitemOpen
		\bibfield  {author} {\bibinfo {author} {\bibfnamefont {J.}~\bibnamefont
				{Casanova}}, \bibinfo {author} {\bibfnamefont {R.}~\bibnamefont {Puebla}},
			\bibinfo {author} {\bibfnamefont {H.}~\bibnamefont {{Moya-Cessa}}},\ and\
			\bibinfo {author} {\bibfnamefont {M.~B.}\ \bibnamefont {Plenio}},\ }\bibfield
		{title} {\bibinfo {title} {Connecting nth order generalised quantum {Rabi}
				models: Emergence of nonlinear spin-boson coupling via spin rotations},\
		}\href {https://doi.org/10.1038/s41534-018-0096-9} {\bibfield  {journal}
			{\bibinfo  {journal} {npj Quantum Inf.}\ }\textbf {\bibinfo {volume} {4}},\
			\bibinfo {pages} {47} (\bibinfo {year} {2018})}\BibitemShut {NoStop}%
		\bibitem [{\citenamefont {Puebla}\ \emph {et~al.}(2019)\citenamefont {Puebla},
			\citenamefont {Casanova}, \citenamefont {Houhou}, \citenamefont {Solano},\
			and\ \citenamefont {Paternostro}}]{pueblaQuantumSimulationMultiphoton2019}%
		\BibitemOpen
		\bibfield  {author} {\bibinfo {author} {\bibfnamefont {R.}~\bibnamefont
				{Puebla}}, \bibinfo {author} {\bibfnamefont {J.}~\bibnamefont {Casanova}},
			\bibinfo {author} {\bibfnamefont {O.}~\bibnamefont {Houhou}}, \bibinfo
			{author} {\bibfnamefont {E.}~\bibnamefont {Solano}},\ and\ \bibinfo {author}
			{\bibfnamefont {M.}~\bibnamefont {Paternostro}},\ }\bibfield  {title}
		{\bibinfo {title} {Quantum simulation of multiphoton and nonlinear
				dissipative spin-boson models},\ }\href
		{https://doi.org/10.1103/PhysRevA.99.032303} {\bibfield  {journal} {\bibinfo
				{journal} {Phys. Rev. A}\ }\textbf {\bibinfo {volume} {99}},\ \bibinfo
			{pages} {032303} (\bibinfo {year} {2019})}\BibitemShut {NoStop}%
		\bibitem [{\citenamefont {Felicetti}\ \emph
			{et~al.}(2018{\natexlab{a}})\citenamefont {Felicetti}, \citenamefont
			{Rossatto}, \citenamefont {Rico}, \citenamefont {Solano},\ and\ \citenamefont
			{{Forn-D{\'i}az}}}]{felicettiTwophotonQuantumRabi2018}%
		\BibitemOpen
		\bibfield  {author} {\bibinfo {author} {\bibfnamefont {S.}~\bibnamefont
				{Felicetti}}, \bibinfo {author} {\bibfnamefont {D.~Z.}\ \bibnamefont
				{Rossatto}}, \bibinfo {author} {\bibfnamefont {E.}~\bibnamefont {Rico}},
			\bibinfo {author} {\bibfnamefont {E.}~\bibnamefont {Solano}},\ and\ \bibinfo
			{author} {\bibfnamefont {P.}~\bibnamefont {{Forn-D{\'i}az}}},\ }\bibfield
		{title} {\bibinfo {title} {Two-photon quantum {Rabi} model with
				superconducting circuits},\ }\href
		{https://doi.org/10.1103/PhysRevA.97.013851} {\bibfield  {journal} {\bibinfo
				{journal} {Phys. Rev. A}\ }\textbf {\bibinfo {volume} {97}},\ \bibinfo
			{pages} {013851} (\bibinfo {year} {2018}{\natexlab{a}})}\BibitemShut
		{NoStop}%
		\bibitem [{\citenamefont {Felicetti}\ \emph
			{et~al.}(2018{\natexlab{b}})\citenamefont {Felicetti}, \citenamefont
			{Hwang},\ and\ \citenamefont
			{Le~Boit{\'e}}}]{felicettiUltrastrongcouplingRegimeNondipolar2018}%
		\BibitemOpen
		\bibfield  {author} {\bibinfo {author} {\bibfnamefont {S.}~\bibnamefont
				{Felicetti}}, \bibinfo {author} {\bibfnamefont {M.-J.}\ \bibnamefont
				{Hwang}},\ and\ \bibinfo {author} {\bibfnamefont {A.}~\bibnamefont
				{Le~Boit{\'e}}},\ }\bibfield  {title} {\bibinfo {title} {Ultrastrong-coupling
				regime of nondipolar light-matter interactions},\ }\href
		{https://doi.org/10.1103/PhysRevA.98.053859} {\bibfield  {journal} {\bibinfo
				{journal} {Phys. Rev. A}\ }\textbf {\bibinfo {volume} {98}},\ \bibinfo
			{pages} {053859} (\bibinfo {year} {2018}{\natexlab{b}})}\BibitemShut
		{NoStop}%
		\bibitem [{\citenamefont {Wang}\ \emph {et~al.}(2025)\citenamefont {Wang},
			\citenamefont {Mercurio}, \citenamefont {Ridolfo}, \citenamefont {Wang},
			\citenamefont {Chen}, \citenamefont {Wang}, \citenamefont {Liu},
			\citenamefont {Sun}, \citenamefont {Li}, \citenamefont {Nori}, \citenamefont
			{Savasta},\ and\ \citenamefont {You}}]{wangStrongCouplingSinglephoton2025}%
		\BibitemOpen
		\bibfield  {author} {\bibinfo {author} {\bibfnamefont {S.-P.}\ \bibnamefont
				{Wang}}, \bibinfo {author} {\bibfnamefont {A.}~\bibnamefont {Mercurio}},
			\bibinfo {author} {\bibfnamefont {A.}~\bibnamefont {Ridolfo}}, \bibinfo
			{author} {\bibfnamefont {Y.}~\bibnamefont {Wang}}, \bibinfo {author}
			{\bibfnamefont {M.}~\bibnamefont {Chen}}, \bibinfo {author} {\bibfnamefont
				{W.}~\bibnamefont {Wang}}, \bibinfo {author} {\bibfnamefont {Y.}~\bibnamefont
				{Liu}}, \bibinfo {author} {\bibfnamefont {H.}~\bibnamefont {Sun}}, \bibinfo
			{author} {\bibfnamefont {T.}~\bibnamefont {Li}}, \bibinfo {author}
			{\bibfnamefont {F.}~\bibnamefont {Nori}}, \bibinfo {author} {\bibfnamefont
				{S.}~\bibnamefont {Savasta}},\ and\ \bibinfo {author} {\bibfnamefont {J.~Q.}\
				\bibnamefont {You}},\ }\bibfield  {title} {\bibinfo {title} {Strong coupling
				between a single-photon and a two-photon {Fock} state},\ }\href
		{https://doi.org/10.1038/s41467-025-63783-7} {\bibfield  {journal} {\bibinfo
				{journal} {Nat. Commun.}\ }\textbf {\bibinfo {volume} {16}},\ \bibinfo
			{pages} {8730} (\bibinfo {year} {2025})}\BibitemShut {NoStop}%
		\bibitem [{\citenamefont {Duan}\ \emph {et~al.}(2016)\citenamefont {Duan},
			\citenamefont {Xie}, \citenamefont {Braak},\ and\ \citenamefont
			{Chen}}]{duanTwophotonRabiModel2016}%
		\BibitemOpen
		\bibfield  {author} {\bibinfo {author} {\bibfnamefont {L.}~\bibnamefont
				{Duan}}, \bibinfo {author} {\bibfnamefont {Y.-F.}\ \bibnamefont {Xie}},
			\bibinfo {author} {\bibfnamefont {D.}~\bibnamefont {Braak}},\ and\ \bibinfo
			{author} {\bibfnamefont {Q.-H.}\ \bibnamefont {Chen}},\ }\bibfield  {title}
		{\bibinfo {title} {Two-photon {Rabi} model: Analytic solutions and spectral
				collapse},\ }\href {https://doi.org/10.1088/1751-8113/49/46/464002}
		{\bibfield  {journal} {\bibinfo  {journal} {J. Phys. A: Math. Theor.}\
			}\textbf {\bibinfo {volume} {49}},\ \bibinfo {pages} {464002} (\bibinfo
			{year} {2016})}\BibitemShut {NoStop}%
		\bibitem [{\citenamefont {Rico}\ \emph {et~al.}(2020)\citenamefont {Rico},
			\citenamefont {{Maldonado-Villamizar}},\ and\ \citenamefont
			{{Rodriguez-Lara}}}]{ricoSpectralCollapseTwophoton2020}%
		\BibitemOpen
		\bibfield  {author} {\bibinfo {author} {\bibfnamefont {R.~J.~A.}\
				\bibnamefont {Rico}}, \bibinfo {author} {\bibfnamefont {F.~H.}\ \bibnamefont
				{{Maldonado-Villamizar}}},\ and\ \bibinfo {author} {\bibfnamefont {B.~M.}\
				\bibnamefont {{Rodriguez-Lara}}},\ }\bibfield  {title} {\bibinfo {title}
			{Spectral collapse in the two-photon quantum {Rabi} model},\ }\href
		{https://doi.org/10.1103/PhysRevA.101.063825} {\bibfield  {journal} {\bibinfo
				{journal} {Phys. Rev. A}\ }\textbf {\bibinfo {volume} {101}},\ \bibinfo
			{pages} {063825} (\bibinfo {year} {2020})}\BibitemShut {NoStop}%
		\bibitem [{\citenamefont {Chan}(2020)}]{chanBoundStatesTwophoton2020}%
		\BibitemOpen
		\bibfield  {author} {\bibinfo {author} {\bibfnamefont {C.~K.}\ \bibnamefont
				{Chan}},\ }\bibfield  {title} {\bibinfo {title} {Bound states of two-photon
				{Rabi} model at the collapse point},\ }\href
		{https://doi.org/10.1088/1751-8121/aba3e0} {\bibfield  {journal} {\bibinfo
				{journal} {J. Phys. A: Math. Theor.}\ }\textbf {\bibinfo {volume} {53}},\
			\bibinfo {pages} {385303} (\bibinfo {year} {2020})}\BibitemShut {NoStop}%
		\bibitem [{\citenamefont
			{Braak}(2023)}]{braakSpectralDeterminantTwophoton2023}%
		\BibitemOpen
		\bibfield  {author} {\bibinfo {author} {\bibfnamefont {D.}~\bibnamefont
				{Braak}},\ }\bibfield  {title} {\bibinfo {title} {Spectral determinant of the
				two-photon quantum {Rabi} model},\ }\href
		{https://doi.org/10.1002/andp.202200519} {\bibfield  {journal} {\bibinfo
				{journal} {Ann. Phys.}\ }\textbf {\bibinfo {volume} {535}},\ \bibinfo {pages}
			{2200519} (\bibinfo {year} {2023})}\BibitemShut {NoStop}%
		\bibitem [{\citenamefont {Li}\ \emph {et~al.}(2025)\citenamefont {Li},
			\citenamefont {Braak},\ and\ \citenamefont
			{Chen}}]{liCriticalSpectrumAnisotropic2025}%
		\BibitemOpen
		\bibfield  {author} {\bibinfo {author} {\bibfnamefont {J.}~\bibnamefont
				{Li}}, \bibinfo {author} {\bibfnamefont {D.}~\bibnamefont {Braak}},\ and\
			\bibinfo {author} {\bibfnamefont {Q.-H.}\ \bibnamefont {Chen}},\ }\bibfield
		{title} {\bibinfo {title} {Critical spectrum of the anisotropic two-photon
				quantum {Rabi} model},\ }\href {https://doi.org/10.1103/PhysRevA.111.043706}
		{\bibfield  {journal} {\bibinfo  {journal} {Phys. Rev. A}\ }\textbf {\bibinfo
				{volume} {111}},\ \bibinfo {pages} {043706} (\bibinfo {year}
			{2025})}\BibitemShut {NoStop}%
		\bibitem [{\citenamefont {Garbe}\ \emph {et~al.}(2017)\citenamefont {Garbe},
			\citenamefont {Egusquiza}, \citenamefont {Solano}, \citenamefont {Ciuti},
			\citenamefont {Coudreau}, \citenamefont {Milman},\ and\ \citenamefont
			{Felicetti}}]{garbeSuperradiantPhaseTransition2017}%
		\BibitemOpen
		\bibfield  {author} {\bibinfo {author} {\bibfnamefont {L.}~\bibnamefont
				{Garbe}}, \bibinfo {author} {\bibfnamefont {I.~L.}\ \bibnamefont
				{Egusquiza}}, \bibinfo {author} {\bibfnamefont {E.}~\bibnamefont {Solano}},
			\bibinfo {author} {\bibfnamefont {C.}~\bibnamefont {Ciuti}}, \bibinfo
			{author} {\bibfnamefont {T.}~\bibnamefont {Coudreau}}, \bibinfo {author}
			{\bibfnamefont {P.}~\bibnamefont {Milman}},\ and\ \bibinfo {author}
			{\bibfnamefont {S.}~\bibnamefont {Felicetti}},\ }\bibfield  {title} {\bibinfo
			{title} {Superradiant phase transition in the ultrastrong-coupling regime of
				the two-photon {Dicke} model},\ }\href
		{https://doi.org/10.1103/PhysRevA.95.053854} {\bibfield  {journal} {\bibinfo
				{journal} {Phys. Rev. A}\ }\textbf {\bibinfo {volume} {95}},\ \bibinfo
			{pages} {053854} (\bibinfo {year} {2017})}\BibitemShut {NoStop}%
		\bibitem [{\citenamefont {Chen}\ and\ \citenamefont
			{Zhang}(2018)}]{chenFinitesizeScalingAnalysis2018}%
		\BibitemOpen
		\bibfield  {author} {\bibinfo {author} {\bibfnamefont {X.-Y.}\ \bibnamefont
				{Chen}}\ and\ \bibinfo {author} {\bibfnamefont {Y.-Y.}\ \bibnamefont
				{Zhang}},\ }\bibfield  {title} {\bibinfo {title} {Finite-size scaling
				analysis in the two-photon {Dicke} model},\ }\href
		{https://doi.org/10.1103/PhysRevA.97.053821} {\bibfield  {journal} {\bibinfo
				{journal} {Phys. Rev. A}\ }\textbf {\bibinfo {volume} {97}},\ \bibinfo
			{pages} {053821} (\bibinfo {year} {2018})}\BibitemShut {NoStop}%
		\bibitem [{\citenamefont {Garbe}\ \emph
			{et~al.}(2020{\natexlab{b}})\citenamefont {Garbe}, \citenamefont {Wade},
			\citenamefont {Minganti}, \citenamefont {Shammah}, \citenamefont
			{Felicetti},\ and\ \citenamefont
			{Nori}}]{garbeDissipationinducedBistabilityTwophoton2020}%
		\BibitemOpen
		\bibfield  {author} {\bibinfo {author} {\bibfnamefont {L.}~\bibnamefont
				{Garbe}}, \bibinfo {author} {\bibfnamefont {P.}~\bibnamefont {Wade}},
			\bibinfo {author} {\bibfnamefont {F.}~\bibnamefont {Minganti}}, \bibinfo
			{author} {\bibfnamefont {N.}~\bibnamefont {Shammah}}, \bibinfo {author}
			{\bibfnamefont {S.}~\bibnamefont {Felicetti}},\ and\ \bibinfo {author}
			{\bibfnamefont {F.}~\bibnamefont {Nori}},\ }\bibfield  {title} {\bibinfo
			{title} {Dissipation-induced bistability in the two-photon {Dicke} model},\
		}\href {https://doi.org/10.1038/s41598-020-69704-6} {\bibfield  {journal}
			{\bibinfo  {journal} {Sci. Rep.}\ }\textbf {\bibinfo {volume} {10}},\
			\bibinfo {pages} {13408} (\bibinfo {year} {2020}{\natexlab{b}})}\BibitemShut
		{NoStop}%
		\bibitem [{\citenamefont {Ram{\'i}rez}\ \emph {et~al.}(2026)\citenamefont
			{Ram{\'i}rez}, \citenamefont {Villase{\~n}or}, \citenamefont {V{\'a}zquez},\
			and\ \citenamefont {Hirsch}}]{ramirezSuperradiantPhaseFinite2026}%
		\BibitemOpen
		\bibfield  {author} {\bibinfo {author} {\bibfnamefont {F.}~\bibnamefont
				{Ram{\'i}rez}}, \bibinfo {author} {\bibfnamefont {D.}~\bibnamefont
				{Villase{\~n}or}}, \bibinfo {author} {\bibfnamefont {N.}~\bibnamefont
				{V{\'a}zquez}},\ and\ \bibinfo {author} {\bibfnamefont {J.~G.}\ \bibnamefont
				{Hirsch}},\ }\bibfield  {title} {\bibinfo {title} {Superradiant phase is a
				finite size effect in two-photon processes},\ }\href
		{https://doi.org/10.1103/syq9-z93n} {\bibfield  {journal} {\bibinfo
				{journal} {Phys. Rev. Lett.}\ }\textbf {\bibinfo {volume} {137}},\ \bibinfo
			{pages} {053603} (\bibinfo {year} {2026})}\BibitemShut {NoStop}%
		\bibitem [{\citenamefont {Li}\ \emph {et~al.}(2026)\citenamefont {Li},
			\citenamefont {Wang}, \citenamefont {Chen},\ and\ \citenamefont
			{Lin}}]{liQuantumCriticalitySpectral2026}%
		\BibitemOpen
		\bibfield  {author} {\bibinfo {author} {\bibfnamefont {J.}~\bibnamefont
				{Li}}, \bibinfo {author} {\bibfnamefont {J.-L.}\ \bibnamefont {Wang}},
			\bibinfo {author} {\bibfnamefont {Q.-H.}\ \bibnamefont {Chen}},\ and\
			\bibinfo {author} {\bibfnamefont {H.-Q.}\ \bibnamefont {Lin}},\ }\bibfield
		{title} {\bibinfo {title} {Quantum criticality from spectral collapse in the
				two-photon {Rabi} model},\ }\href {https://doi.org/10.1103/ctp9-k77q}
		{\bibfield  {journal} {\bibinfo  {journal} {Phys. Rev. Lett.}\ }\textbf
			{\bibinfo {volume} {137}},\ \bibinfo {pages} {123602} (\bibinfo {year}
			{2026})}\BibitemShut {NoStop}%
		\bibitem [{sup()}]{supplement}%
		\BibitemOpen
		\href@noop {} {}\bibinfo {note} {See Supplemental Material for further
			explanation and details of the calculation.}\BibitemShut {Stop}%
		\bibitem [{\citenamefont {Baumann}\ \emph {et~al.}(2011)\citenamefont
			{Baumann}, \citenamefont {Mottl}, \citenamefont {Brennecke},\ and\
			\citenamefont {Esslinger}}]{baumannExploringSymmetryBreaking2011}%
		\BibitemOpen
		\bibfield  {author} {\bibinfo {author} {\bibfnamefont {K.}~\bibnamefont
				{Baumann}}, \bibinfo {author} {\bibfnamefont {R.}~\bibnamefont {Mottl}},
			\bibinfo {author} {\bibfnamefont {F.}~\bibnamefont {Brennecke}},\ and\
			\bibinfo {author} {\bibfnamefont {T.}~\bibnamefont {Esslinger}},\ }\bibfield
		{title} {\bibinfo {title} {Exploring symmetry breaking at the {Dicke} quantum
				phase transition},\ }\href {https://doi.org/10.1103/PhysRevLett.107.140402}
		{\bibfield  {journal} {\bibinfo  {journal} {Phys. Rev. Lett.}\ }\textbf
			{\bibinfo {volume} {107}},\ \bibinfo {pages} {140402} (\bibinfo {year}
			{2011})}\BibitemShut {NoStop}%
	\end{thebibliography}
	%
	
\end{document}


\title{Fates of Exact Finite-$N$ First-Order Transitions: Coherent Suppression versus Spectral-Collapse Termination\\
		-Supplemental Material-}
	
	\author{Jiong Li$^1$}
	\author{Li-Wei Duan$^2$}
	\author{Qing-Hu Chen$^{3,1,}$}
	\email{qhchen@zju.edu.cn}
	\author{Hai-Qing Lin$^{1,}$}
	\email{hqlin@zju.edu.cn}
	
	\affiliation{
		$^1$ Institute for Advanced Study in Physics and School of Physics, Zhejiang University, Hangzhou 310058, China \\
		$^2$ Department of Physics, Zhejiang Normal University, Jinhua 321004, China\\
		$^3$ Zhejiang Key Laboratory of Micro-Nano Quantum Chips and Quantum Control, School of Physics,
		Zhejiang University, Hangzhou 310058, China
	}
	
	\date{\today}
	
	\maketitle
	
	\section{EXACT FINITE-$N$ JUDDIAN STATES AND CLOSURE CONDITION}
	\label{sec:supp_truncation}
	
	We derive the lowest finite-$N$ Juddian solution directly from Gaussian boundary closure, without invoking the two-channel determinant representation developed in Sec.~\ref{sec:supp_determinant}. A spin-space similarity transformation, followed by displacement for $k=1$ or squeezing for $k=2$, triangularizes the highest-weight boundary in the corresponding Gaussian representation. The resulting closure conditions determine the Juddian coupling and yield an explicit nonzero eigenstate for each finite-$N$ Hamiltonian.
	
	Throughout the Supplemental Material, we work in the fully symmetric spin sector $j=N/2$ and assume $0<r<1$. The limiting cases $r=0$ and $r=1$ require separate treatment and are not considered here.	

	\subsection{Highest-weight Gaussian boundary triangularization}
	\label{sec:supp_gaussian_triangularization}
	
	We begin with the unified Hamiltonian for the one- and two-photon anisotropic Dicke models (ADMs),
	\begin{equation}
		H_k = \omega a^\dagger a + \Delta J_z + \frac{g}{\sqrt{N}} \left( a^k J_+ + a^{\dagger k} J_- \right) + \frac{rg}{\sqrt{N}} \left( a^kJ_- + a^{\dagger k}J_+ \right),
		\quad k=1,2.
		\label{eq:supp_Hk}
	\end{equation}
	We construct transformations that eliminate the boson-lowering terms from the highest-weight spin projection onto $\ket{j,j}$.
	
	\subsubsection{One-photon model}
	
	For $k=1$, apply the invertible spin-space similarity transformation
	\begin{equation}
		H_{s,0}^{(1)} = P_1 H_1 P_1^{-1}, \quad
		P_1 = \bigotimes_{n=1}^{N} P_1^{(n)},
		\label{eq:supp_P1_def}
	\end{equation}
	with
	\begin{equation}
		P_1^{(n)} = \frac{1 + \sqrt{r}}{2 \sqrt{2}} + \frac{\sqrt{r} - 1}{2\sqrt{2}} \sigma_z^{(n)} + \frac{1 - \sqrt{r}}{2 \sqrt{2}} \sigma_x^{(n)} + \frac{1 + \sqrt{r}}{2 \sqrt{2}} i \sigma_y^{(n)} .
		\label{eq:supp_P1}
	\end{equation}
	For $r>0$, $P_1$ is invertible, and hence the similarity transformation preserves the spectrum. The transformed Hamiltonian is
	\begin{equation}
		H_{s,0}^{(1)} = \omega a^\dagger a - \Delta J_x + \frac{2g \sqrt{r}}{\sqrt{N}}(a + a^\dagger) J_z + \frac{g (1-r)^2}{\sqrt{Nr}} a^\dagger J_z - \frac{g (1-r^2)}{\sqrt{Nr}} a^\dagger iJ_y .
		\label{eq:supp_H10}
	\end{equation}
	
	Next apply the displacement
	\begin{equation}
		D(\alpha) = \exp \! \left[ \alpha (a^\dagger - a) \right], \quad
		\alpha = \frac{\sqrt{Nr} g}{\omega},
		\label{eq:supp_D_alpha}
	\end{equation}
	and define
	\begin{equation}
		H_s^{(1)} = D(\alpha) H_{s,0}^{(1)} D(-\alpha).
		\label{eq:supp_H1s_def}
	\end{equation}
	Using $D(\alpha) a D(-\alpha)=a-\alpha$, we obtain
	\begin{equation}
		H_s^{(1)} = \omega a^\dagger a - \omega \alpha(a + a^\dagger) \left( 1 - \frac{J_z}{j} \right) + \omega \alpha^2 \left[ 1 - \frac{(1+r)^2}{2r} \frac{J_z}{j} \right] + \omega \frac{\alpha}{j} \frac{(1-r)^2}{2r} a^\dagger J_z - \Delta J_x - \omega \frac{\alpha}{j} \frac{1-r^2}{2r} (a^\dagger - \alpha) iJ_y .
		\label{eq:supp_H1s}
	\end{equation}
	Since
	\begin{equation}
		\left(1 - \frac{J_z}{j} \right) \ket{j,j}=0,
		\label{eq:supp_highest_weight_1}
	\end{equation}
	the only boson-number-lowering term proportional to $a$ vanishes in the highest-weight projection. The remaining boson-number-changing terms therefore contain only $a^\dagger$, making the boundary triangular in the displaced representation.
	
	\subsubsection{Two-photon model}
	
	For $k=2$, apply the invertible spin-space similarity transformation
	\begin{equation}
		H_{s,0}^{(2)} = P_2 H_2 P_2^{-1}, \quad
		P_2 = \bigotimes_{n=1}^{N}P_2^{(n)},
		\label{eq:supp_P2_def}
	\end{equation}
	where
	\begin{equation}
		P_2^{(n)} = \frac{\sqrt{r_2} + \sqrt{r_1}}{2\sqrt{2r_1}} + \frac{\sqrt{r_2} - \sqrt{r_1}}{2\sqrt{2r_1}} \sigma_z^{(n)} + \frac{\sqrt{r_1} - \sqrt{r_2}}{2\sqrt{2r_1}} \sigma_x^{(n)} + \frac{\sqrt{r_2} + \sqrt{r_1}}{2 \sqrt{2r_1}} i\sigma_y^{(n)}.
		\label{eq:supp_P2}
	\end{equation}
	Here $r_1,r_2>0$ are auxiliary parameters specified below. For positive $r_1$ and $r_2$, $P_2$ is invertible. The collective-spin operators transform as
	\begin{equation}
		P_2 J_+ P_2^{-1} = \sqrt{\frac{r_2}{r_1}} (J_z + iJ_y), \quad
		P_2 J_- P_2^{-1} = \sqrt{\frac{r_1}{r_2}} (J_z - iJ_y), \quad
		P_2 J_z P_2^{-1} = -J_x,
		\label{eq:supp_P2_spin}
	\end{equation}
	which gives
	\begin{equation}
		H_{s,0}^{(2)} = \omega a^\dagger a - \Delta J_x + \frac{g}{\sqrt{N}} \left[ \sqrt{\frac{r_1}{r_2}} (ra^2 + a^{\dagger2}) + \sqrt{\frac{r_2}{r_1}} (a^2 + ra^{\dagger2}) \right] J_z + \frac{g}{\sqrt{N}} \left[ \sqrt{\frac{r_2}{r_1}} (a^2 + ra^{\dagger2}) - \sqrt{\frac{r_1}{r_2}} (ra^2 + a^{\dagger2}) \right] iJ_y.
		\label{eq:supp_H20}
	\end{equation}
	
	Introduce the squeezing transformation
	\begin{equation}
		S(\theta) = \exp \! \left[ \frac{\theta}{2} (a^{\dagger2} - a^2) \right],
		\label{eq:supp_S_theta}
	\end{equation}
	with
	\begin{equation}
		\theta = \frac{1}{2} \ln \! \left[ \frac{\beta_- + \sqrt{\beta_-^2 - \beta_+^2}}{\beta_+} \right], \quad
		\beta_\pm = \sqrt{ 1 - \frac{Ng^2}{\omega^2}(1 \pm r)^2 },
		\label{eq:supp_theta_beta}
	\end{equation}
	and choose
	\begin{equation}
		r_1 = \cosh^2\theta  +r\sinh^2\theta, \quad
		r_2 = r\cosh^2\theta + \sinh^2\theta.
		\label{eq:supp_r1r2}
	\end{equation}
	These definitions imply
	\begin{equation}
		\beta_-^2 - \beta_+^2 = \frac{4Nr g^2}{\omega^2}, \quad 
		r_1 r_2 = \frac{r}{\beta_+^2}, \quad
		r_1 - r_2 = 1-r, \quad 
		r_1 + r_2 = (1+r) \frac{\beta_-}{\beta_+}.
		\label{eq:supp_r_beta_identities}
	\end{equation}
	In the domain $g>0$ and $\beta_+>0$ considered below, $0<r<1$ implies $\beta_- > 0$ and $\beta_-^2 - \beta_+^2 > 0$; hence $\theta$ is real.
	
	Defining
	\begin{equation}
		H_s^{(2)} = S(\theta) H_{s,0}^{(2)} S^{-1}(\theta),
		\label{eq:supp_H2s_def}
	\end{equation}
	we obtain
	\begin{align}
		H_s^{(2)} ={}& \omega \cosh2\theta \left( a^\dagger a + \frac{1}{2} \right)\left[ 1 - \frac{2g^2}{\omega^2} (1+r)^2 J_z \right] - \frac{\omega}{2} + \frac{\omega \sinh2\theta}{N} (J_z-j) (a^2 + a^{\dagger2}) + \frac{g (1-r)^2 \beta_+ }{\sqrt{Nr}} a^{\dagger2} J_z \nonumber\\
		&- \Delta J_x - \frac{g (1-r^2) \beta_-}{\sqrt{Nr}} \left[ a^{\dagger2} - \left(a^\dagger a + \frac{1}{2} \right) \tanh2\theta \right] iJ_y ,
		\label{eq:supp_H2s}
	\end{align}
	where
	\begin{equation}
		\sinh2\theta = \frac{2 \sqrt{Nr} g}{\omega \beta_+}, \quad
		\cosh2\theta = \frac{\beta_-}{\beta_+}.
		\label{eq:supp_theta_relations}
	\end{equation}
	Since
	\begin{equation}
		(J_z-j) \ket{j,j} = 0,
		\label{eq:supp_highest_weight_2}
	\end{equation}
	the only pair-lowering term proportional to $a^2$ vanishes in the highest-weight projection. The remaining boson-number-changing terms therefore contain only $a^{\dagger2}$, making the boundary triangular within each even- or odd-boson subspace.
	
	\subsection{Gaussian-vacuum closure and the Juddian condition}
	\label{sec:supp_lowest_pole}
	
	We expand an eigenstate of the transformed Hamiltonian as
	\begin{equation}
		\ket{\Phi_k} = \sum_{\nu=0}^{\infty} \sum_{m=-j}^{j} c_{\nu,m}^{(k)} \ket{k\nu;j,m}, \quad
		\ket{k\nu;j,m} \equiv \ket{k\nu} \otimes \ket{j,m}.
		\label{eq:supp_expansion}
	\end{equation}
	For $k=1$, $\ket{k\nu}$ labels oscillator basis states in the displaced representation. For $k=2$, it labels the even-boson basis states in the squeezed representation.
	
	Projecting $H_s^{(k)} \ket{\Phi_k} = E \ket{\Phi_k}$ onto the highest-weight Gaussian vacuum $\ket{0;j,j}$ gives
	\begin{equation}
		\left( E - E_{\mathrm{pole}}^{(k)} \right) c_{0,j}^{(k)} = R_{0,j}^{(k)},
		\label{eq:supp_boundary_equation}
	\end{equation}
	where
	\begin{equation}
		R_{0,j}^{(k)} = \sum_{(\nu,m) \neq (0,j)} \bra{0;j,j} H_s^{(k)} \ket{k\nu;j,m} c_{\nu,m}^{(k)} .
		\label{eq:supp_R_boundary}
	\end{equation}
	The diagonal highest-weight vacuum matrix elements are
	\begin{align}
		E_{\mathrm{pole}}^{(1)} &= -\omega \frac{1+r^2}{2r} \alpha^2,
		\label{eq:supp_Epole1} \\
		E_{\mathrm{pole}}^{(2)} &= \frac{\omega}{2} (\beta_+\beta_- -1).
		\label{eq:supp_Epole2}
	\end{align}
	We refer to the energies in Eqs.~\eqref{eq:supp_Epole1} and \eqref{eq:supp_Epole2} as the lowest-pole energies because solving Eq.~\eqref{eq:supp_boundary_equation} for $c_{0,j}^{(k)}$ introduces the denominator $E-E_{\mathrm{pole}}^{(k)}$. 
	
	We now seek the lowest closed solution at $E = E_{\mathrm{pole}}^{(k)}$ within the Gaussian-vacuum sector,
	\begin{equation}
		\ket{\Phi_k} = \ket{0} \otimes \ket{\chi_j^{(k)}}, \quad
		\ket{\chi_j^{(k)}} = \sum_{m=-j}^{j} c_{0,m}^{(k)} \ket{j,m}.
		\label{eq:supp_vacuum_ansatz}
	\end{equation}
	From Eqs.~\eqref{eq:supp_H1s} and \eqref{eq:supp_H2s}, the action of the transformed Hamiltonian on the Gaussian-vacuum sector has the form
	\begin{equation}
		H_s^{(k)} \ket{\Phi_k} = \ket{0} \otimes H_{\mathrm{spin}}^{(k)} \ket{\chi_j^{(k)}} + a^{\dagger k} \ket{0} \otimes K_k \ket{\chi_j^{(k)}}.
		\label{eq:supp_vacuum_action}
	\end{equation}
	Thus the vacuum sector closes exactly if
	\begin{equation}
		K_k \ket{\chi_j^{(k)}}=0,
		\label{eq:supp_kernel_condition}
	\end{equation}
	and the remaining vacuum-sector equation is
	\begin{equation}
		H_{\mathrm{spin}}^{(k)} \ket{\chi_j^{(k)}} = E_J^{(k)} \ket{\chi_j^{(k)}}.
		\label{eq:supp_projected_eigen}
	\end{equation}
	A nonzero spin state satisfying Eqs.~\eqref{eq:supp_kernel_condition} and \eqref{eq:supp_projected_eigen} is constructed explicitly below.
	
	At $E = E_{\mathrm{pole}}^{(k)}$, Eq.~\eqref{eq:supp_boundary_equation} reduces to the highest-weight boundary condition
	\begin{equation}
		R_{0,j}^{(k)} = 0.
		\label{eq:supp_pole_regularization}
	\end{equation} 
	Within the vacuum ansatz, the collective-spin dependence of $H_s^{(k)}$ is at most linear in the ladder operators. Using
	\begin{equation}
		J_x = \frac{J_+ + J_-}{2}, \quad
		iJ_y = \frac{J_+ - J_-}{2},
		\label{eq:supp_spin_ladder_relations}
	\end{equation}
	the only off-diagonal vacuum-sector contribution to Eq.~\eqref{eq:supp_pole_regularization} connects $\ket{j,j-1}$ to $\ket{j,j}$. Direct evaluation of Eqs.~\eqref{eq:supp_H1s} and \eqref{eq:supp_H2s} gives, for both $k=1$ and $k=2$,
	\begin{equation}
		0 = R_{0,j}^{(k)} = \bra{0;j,j} H_s^{(k)} \ket{0;j,j-1} c_{0,j-1}^{(k)} = \left[ \frac{(1-r^2) g^2}{2\omega} - \frac{\Delta}{2} \right] \bra{j,j} J_+ \ket{j,j-1} c_{0,j-1}^{(k)}.
		\label{eq:supp_common_boundary}
	\end{equation}
	For the lowest closed branch constructed in Sec.~\ref{sec:supp_exact_closure}, $c_{0,j-1}^{(k)} \neq 0$. Moreover,
	\begin{equation}
		\bra{j,j} J_+ \ket{j,j-1} = \sqrt{2j} \neq 0.
	\end{equation}
	The highest-weight boundary condition therefore yields
	\begin{equation}
		g_J^2 (1 - r^2) = \Delta \omega, \quad 
		g_J = \sqrt{\frac{\Delta \omega}{1 - r^2}}.
		\label{eq:supp_Juddian_condition}
	\end{equation}
	
	Equation~\eqref{eq:supp_Juddian_condition} is independent of $j=N/2$ and follows directly from finite-$N$ Gaussian-vacuum closure. Neither the normal-state instability thresholds nor the coupling-matrix determinant has been used in its derivation. The equivalent two-channel determinant representation is established independently in Sec.~\ref{sec:supp_determinant}.
	
	\subsection{Explicit Gaussian product eigenstates}
	\label{sec:supp_exact_closure}
	
	We complete the Gaussian-closure construction by explicitly solving Eqs.~\eqref{eq:supp_kernel_condition} and \eqref{eq:supp_projected_eigen} at $g=g_J$ and obtaining nonzero eigenstates. We use the normalized collective-spin coherent states
	\begin{equation}
		\ket{\eta}_j = \frac{e^{\eta J_-}}{(1 + |\eta|^2)^j} \ket{j,j},
		\label{eq:supp_spin_coherent}
	\end{equation}
	together with
	\begin{equation}
		e^{-\eta J_-} J_z e^{\eta J_-} = J_z - \eta J_-, \quad
		e^{-\eta J_-} J_+ e^{\eta J_-} = J_+ + 2\eta J_z - \eta^2 J_-.
		\label{eq:supp_spin_identities}
	\end{equation}
	For convenience, define
	\begin{equation}
		\rho \equiv \frac{1-r}{1+r}, \quad 0<\rho<1 .
		\label{eq:supp_rho}
	\end{equation}
	
	\subsubsection{One-photon branch}
	
	At $g=g_J$, Eq.~\eqref{eq:supp_Epole1} gives
	\begin{equation}
		E_J^{(1)} = - \omega \frac{1+r^2}{2r} \alpha^2 = - \frac{N(1+r^2)}{2(1-r^2)} \Delta .
		\label{eq:supp_EJ1}
	\end{equation}
	The coefficient of $a^\dagger \ket{0}$ in $H_s^{(1)}$~\eqref{eq:supp_H1s} defines the leakage operator
	\begin{equation}
		K_1 = \omega \alpha \left[ - 1 + \frac{1+r^2}{2r} \frac{J_z}{j} - \frac{1-r^2}{2r} \frac{iJ_y}{j} \right].
		\label{eq:supp_K1}
	\end{equation}
	Using Eq.~\eqref{eq:supp_spin_identities}, the kernel condition $K_1 \ket{\eta_1}_j = 0$ reduces to
	\begin{align}
		\frac{1+r^2}{2r} -\frac{1-r^2}{2r} \eta_1 - 1 &= 0, \nonumber \\
		\frac{1+r^2}{2r} \eta_1 - \frac{1-r^2}{4r} (1 + \eta_1^2) &= 0.
		\label{eq:supp_eta1_conditions}
	\end{align}
	These equations admit the common solution
	\begin{equation}
		\eta_1 = \rho = \frac{1-r}{1+r}.
		\label{eq:supp_eta1}
	\end{equation}
	
	Projection of $H_s^{(1)}$ onto the displaced vacuum gives
	\begin{equation}
		H_{\mathrm{spin}}^{(1)} = \omega \left[ \alpha^2 - \frac{\alpha^2}{j} \frac{(1+r)^2}{2r} J_z + \frac{\alpha^2}{j} \frac{1-r^2}{2r} iJ_y - \frac{\Delta}{\omega} J_x \right].
		\label{eq:supp_Hspin1}
	\end{equation}
	Substituting Eqs.~\eqref{eq:supp_Juddian_condition} and \eqref{eq:supp_eta1} into Eq.~\eqref{eq:supp_Hspin1}, and using Eq.~\eqref{eq:supp_spin_identities}, gives
	\begin{equation}
		H_{\mathrm{spin}}^{(1)} \ket{\eta_1}_j = E_J^{(1)} \ket{\eta_1}_j .
		\label{eq:supp_Hspin1_eigen}
	\end{equation}
	Hence Eqs.~\eqref{eq:supp_kernel_condition} and \eqref{eq:supp_projected_eigen} are both satisfied, and
	\begin{equation}
		\ket{\Phi_J^{(1)}} = \ket{0} \otimes \ket{\eta_1}_j
		\label{eq:supp_PhiJ1}
	\end{equation}
	is an exact nonzero eigenstate of $H_s^{(1)}$.
	
	Applying the inverse bosonic and spin-space transformations and normalizing the resulting state gives
	\begin{equation}
		\ket{\Psi_L^{(1)}} = \ket{-\alpha} \otimes \ket{\zeta_1}_j, \quad
		\zeta_1 = \frac{1}{\sqrt{r}}, \quad
		\ket{-\alpha}=D(-\alpha)\ket{0},
		\label{eq:supp_PsiL1}
	\end{equation}
	where $\zeta_1$ follows from the action of $P_1^{-1}$ on $\ket{\eta_1}_j$. Therefore,
	\begin{equation}
		H_1 \ket{\Psi_L^{(1)}} = E_J^{(1)} \ket{\Psi_L^{(1)}} .
		\label{eq:supp_exact1}
	\end{equation}
	
	\subsubsection{Two-photon branch}
	
	For $k=2$, Eq.~\eqref{eq:supp_Epole2} gives at $g=g_J$
	\begin{equation}
		E_J^{(2)} = \frac{\omega}{2} (\beta_+ \beta_- -1) = \frac{\omega}{2} \left[ \sqrt{\left( 1 - \frac{N}{\rho} \frac{\Delta}{\omega} \right)\left( 1 - N\rho \frac{\Delta}{\omega} \right)} - 1 \right],
		\label{eq:supp_EJ2}
	\end{equation}
	The coefficient of $a^{\dagger2} \ket{0}$ in $H_s^{(2)}$~\eqref{eq:supp_H2s} defines
	\begin{equation}
		K_2 = \frac{g}{\sqrt{Nr}} \left[ \frac{2r}{\beta_+} (J_z-j) + (1-r)^2 \beta_+ J_z - (1-r^2) \beta_- iJ_y \right].
		\label{eq:supp_K2}
	\end{equation}
	Using Eq.~\eqref{eq:supp_spin_identities}, the kernel condition $K_2 \ket{\eta_2}_j = 0$ reduces to
	\begin{align}
		(1-r)^2 \beta_+ - (1-r^2)\beta_-\eta_2 &= 0, \nonumber \\
		\frac{(1-r^2) \beta_-}{2} (1 + \eta_2^2) - \frac{2r}{\beta_+} \eta_2 - (1-r)^2 \beta_+ \eta_2 &= 0.
		\label{eq:supp_eta2_conditions}
	\end{align}
	At $g=g_J$, their common solution is
	\begin{equation}
		\eta_2 = \rho \frac{\beta_+}{\beta_-}.
		\label{eq:supp_eta2}
	\end{equation}
	
	Projection of $H_s^{(2)}$ onto the squeezed vacuum gives
	\begin{equation}
		H_{\mathrm{spin}}^{(2)} = \frac{\omega}{2} \frac{\beta_-}{\beta_+} \left[ 1 - \frac{2g^2}{\omega^2} (1+r)^2 J_z \right] - \frac{\omega}{2} - \Delta J_x + \frac{(1-r^2) g^2}{\omega} iJ_y .
		\label{eq:supp_Hspin2}
	\end{equation}
	Substituting Eqs.~\eqref{eq:supp_Juddian_condition} and \eqref{eq:supp_eta2} into Eq.~\eqref{eq:supp_Hspin2}, and using Eq.~\eqref{eq:supp_spin_identities}, gives
	\begin{equation}
		H_{\mathrm{spin}}^{(2)} \ket{\eta_2}_j = E_J^{(2)} \ket{\eta_2}_j.
		\label{eq:supp_Hspin2_eigen}
	\end{equation}
	Hence both closure conditions are satisfied, and
	\begin{equation}
		\ket{\Phi_J^{(2)}} = \ket{0} \otimes \ket{\eta_2}_j
		\label{eq:supp_PhiJ2}
	\end{equation}
	is an exact nonzero eigenstate of $H_s^{(2)}$.
	
	Applying the inverse bosonic and spin-space transformations and normalizing the resulting state gives
	\begin{equation}
		\ket{\Psi_L^{(2)}} = \ket{-\theta}_2 \otimes \ket{\zeta_2}_j, \quad 
		\zeta_2=\sqrt{\frac{r_1}{r_2}}, \quad
		\ket{-\theta}_2 = S(-\theta) \ket{0},
		\label{eq:supp_PsiL2}
	\end{equation}
	where $\zeta_2$ follows from the action of $P_2^{-1}$ on $\ket{\eta_2}_j$. Therefore,
	\begin{equation}
		H_2 \ket{\Psi_L^{(2)}} = E_J^{(2)} \ket{\Psi_L^{(2)}}.
		\label{eq:supp_exact2}
	\end{equation}
	
	For the two-photon branch, the squeezed vacuum is normalizable only for a finite real squeezing parameter $\theta$. At $g=g_J$,
	\begin{equation}
		\beta_+^2 = 1 - \frac{N}{\rho} \frac{\Delta}{\omega}, \quad
		\beta_-^2 = 1 - N \rho \frac{\Delta}{\omega}.
		\label{eq:supp_beta_J}
	\end{equation}
	Since $0<\rho<1$, $\beta_+^2 < \beta_-^2$, and the stronger condition is therefore
	\begin{equation}
		\beta_+^2>0 \quad \Longleftrightarrow \quad \frac{N\Delta}{\omega} < \rho = \frac{1-r}{1+r}.
		\label{eq:supp_normalizability}
	\end{equation}
	Using Eq.~\eqref{eq:supp_Juddian_condition}, this is equivalently
	\begin{equation}
		\sqrt N g_J (1+r) < \omega .
		\label{eq:supp_normalizability_g}
	\end{equation}
	Defining
	\begin{equation}
		g_c \equiv \frac{\omega}{\sqrt N(1+r)}, \quad
		\Delta_c \equiv \frac{\omega}{N} \frac{1-r}{1+r},
		\label{eq:supp_gc_Deltac_prelim}
	\end{equation}
	we obtain
	\begin{equation}
		g_J<g_c \quad \Longleftrightarrow \quad \Delta < \Delta_c .
		\label{eq:supp_normalizability_equiv}
	\end{equation}
	The identification of $g_c$ with the spectral-collapse boundary is established in Sec.~\ref{sec:supp_distinct_fates}.
	
	Equations~\eqref{eq:supp_exact1} and \eqref{eq:supp_exact2} therefore establish explicit nonzero Gaussian product eigenstates of the full finite-$N$ Hamiltonians at the common Juddian condition
	\begin{equation}
		g_J^2 (1-r^2) = \Delta\omega,
		\label{eq:supp_common_gJ_final}
	\end{equation}
	with the additional normalizability requirement \eqref{eq:supp_normalizability} for the two-photon branch. This completes the direct Gaussian-closure construction of the lowest exact Juddian branch. The equivalent two-channel determinant representation of Eq.~\eqref{eq:supp_common_gJ_final} is derived independently in Sec.~\ref{sec:supp_determinant}.
	
	\section{TWO-CHANNEL STRUCTURE OF THE LOWEST JUDDIAN CONDITION}
	\label{sec:supp_determinant}
	
	Section~\ref{sec:supp_truncation} derived the lowest finite-$N$ Juddian condition,
	\begin{equation}
		g_J^2 (1-r^2) = \Delta \omega
		\label{eq:supp_gJ_from_truncation}
	\end{equation}
	directly from Gaussian boundary closure, without invoking the coupling-matrix structure. We now show how the channel product $g^2(1-r^2)$, identified below as $\det G$, enters the highest-weight boundary coefficient under the spin-space similarity and canonical bosonic transformations. We subsequently relate the same two-channel structure to the channel-resolved normal-state instability thresholds and to the exact Juddian crossing.
	
	\subsection{Coupling-matrix representation}
	\label{sec:supp_spin_determinant}
	
	The interaction can be written as
	\begin{equation}
		H_{\mathrm{int}}^{(k)} = A_k^T G \mathbf{J}, \quad
		A_k = \frac{1}{\sqrt{N}} \begin{pmatrix}
			a^k \\ a^{\dagger k}
		\end{pmatrix}, \quad
		\mathbf{J} = \begin{pmatrix}
			J_+ \\ J_-
		\end{pmatrix},
		\label{eq:supp_matrix_interaction}
	\end{equation}
	with the coupling matrix
	\begin{equation}
		G = g
		\begin{pmatrix}
			1&r \\ r&1
		\end{pmatrix}.
		\label{eq:supp_G_matrix}
	\end{equation}
	Its eigenvalues are the two channel strengths
	\begin{equation}
		G_\pm = g (1 \pm r),
	\end{equation}
	so that
	\begin{equation}
		\det G = G_+ G_- = g^2 (1-r^2).
		\label{eq:supp_detG}
	\end{equation}
	
	The spin-space similarity transformations introduced in Sec.~\ref{sec:supp_gaussian_triangularization} reorganize the collective-spin operators in the basis
	\begin{equation}
		\mathbf{J}' = \begin{pmatrix}
			J_z \\ iJ_y
		\end{pmatrix}.
	\end{equation}
	Their action can be written in the common form
	\begin{equation}
		\begin{pmatrix}
			P(t) J_+ P^{-1}(t)\\
			P(t) J_- P^{-1}(t)
		\end{pmatrix} = T_s(t) \mathbf{J}', \quad
		T_s(t) =
		\begin{pmatrix}
			t&t \\
			t^{-1} & -t^{-1}
		\end{pmatrix}.
		\label{eq:supp_Ts}
	\end{equation}
	The one- and two-photon transformations correspond respectively to
	\begin{equation}
		t_1 = \sqrt{r}, \quad
		t_2 = \sqrt{\frac{r_2}{r_1}},
		\label{eq:supp_t12}
	\end{equation}
	with $r_1$ and $r_2$ defined in Eq.~\eqref{eq:supp_r1r2}. Since
	\begin{equation}
		\det T_s(t)=-2
	\end{equation}
	is independent of $t$, the interaction block in the $(J_z,iJ_y)$ basis,
	\begin{equation}
		C(t) = GT_s(t),
		\label{eq:supp_Ct}
	\end{equation}
	satisfies
	\begin{equation}
		\det C(t) = \det G \det T_s(t) = -2\det G.
		\label{eq:supp_detGTs}
	\end{equation}
	Equation \eqref{eq:supp_detGTs} shows that the determinant of the spin-transformed interaction block is proportional to $G_+G_-$, with the proportionality factor $-2$ independent of $t$.
	
	The subsequent bosonic transformation triangularizes the boundary-relevant number-changing terms. We collect the corresponding coefficients $u_k$ and $w_k$ into the schematic triangular block
	\begin{equation}
		C_k = \begin{pmatrix}
			u_k & 0\\
			* & w_k
		\end{pmatrix}, \quad
		\det C_k = u_k w_k.
		\label{eq:supp_Gk_triangular}
	\end{equation}
	Here $u_k$ denotes the $J_z$ pivot coefficient selected by the Gaussian triangularization, while $w_k$ denotes the residual $iJ_y$ coefficient entering the highest-weight boundary equation.
	
	For the one-photon construction, the displacement acts directly on the boundary-relevant number-changing terms. For the two-photon construction, the triangular block contains only the boundary-relevant pair-changing coefficients. The complete boundary coefficient must also include the number-preserving contribution generated by the full canonical squeezing transformation. The following two subsections evaluate these two cases separately and show how the same channel product $\det G=G_+G_-$ enters the resulting boundary condition.
	
	\subsection{One-photon boundary contribution}
	\label{sec:supp_det_one_photon}
	
	For $k=1$, the boundary-relevant pivot and residual coefficients introduced in Sec.~\ref{sec:supp_spin_determinant} are 
	\begin{equation}
		u_1 = \sqrt{N} \frac{\omega \alpha}{j}, \quad
		w_1 = - \sqrt{N} \omega\frac{\alpha}{j} \frac{1-r^2}{2r}.
		\label{eq:supp_u1w1}
	\end{equation}
	Using $\alpha = \sqrt{Nr} g/\omega$ and $j=N/2$, their product is
	\begin{equation}
		u_1w_1 = -2g^2(1-r^2) = -2\det G.
		\label{eq:supp_u1w1_det}
	\end{equation}
	Thus the determinant structure of the spin-transformed interaction block is retained by the boundary-relevant coefficients after displacement.
	
	The displacement parameter can equivalently be expressed through the pivot coefficient as
	\begin{equation}
		\alpha = \frac{ju_1}{\sqrt{N} \omega}.
		\label{eq:supp_alpha_u1}
	\end{equation}
	Under displacement, the residual term $(w_1/\sqrt{N}) a^\dagger iJ_y$ acquires the vacuum-preserving contribution $-(w_1\alpha/\sqrt{N}) iJ_y$. Using $iJ_y = (J_+-J_-)/2$, the interaction-generated $J_+$ coefficient in the highest-weight boundary equation is
	\begin{equation}
		\Sigma_{\mathrm{edge}}^{(1)} = - \frac{w_1 \alpha}{2\sqrt{N}} = - \frac{j u_1 w_1}{2N \omega} = \frac{\det G}{2\omega}.
		\label{eq:supp_Sigma1_det}
	\end{equation}
	Therefore, in the one-photon ADM, displacement transfers the channel product $G_+G_-=\det G$ directly into the interaction-generated highest-weight boundary coefficient.

	\subsection{Two-photon boundary contribution}
	\label{sec:supp_det_two_photon}
	
	For $k=2$, the boundary-relevant pair-changing coefficients introduced in Sec.~\ref{sec:supp_spin_determinant} are
	\begin{equation}
		u_2 = \frac{\omega \sinh2\theta}{\sqrt{N}}, \quad
		w_2 = - \frac{g(1-r^2) \beta_-}{\sqrt{r}}.
		\label{eq:supp_u2w2}
	\end{equation}
	Using Eq.~\eqref{eq:supp_theta_relations}, their product is
	\begin{equation}
		u_2 w_2 = -2\cosh2\theta \det G.
		\label{eq:supp_u2w2_det}
	\end{equation}
	Unlike the one-photon case, the pair-changing block therefore contains an additional Bogoliubov factor $\cosh 2\theta$.
	
	To include the number-preserving contribution, we represent squeezing on the full quadratic-operator space spanned by
	\begin{equation}
		\mathbf{v} =
		\begin{pmatrix}
			a^2 &  a^{\dagger2} & 2a^\dagger a+1
		\end{pmatrix}^T.
		\label{eq:supp_X_vector}
	\end{equation}
	Squeezing acts linearly on this operator space,
	\begin{equation}
		S(\theta) \mathbf{v} S^{-1}(\theta) = M(\theta) \mathbf{v},
		\label{eq:supp_M_transform}
	\end{equation}
	where
	\begin{equation}
		M(\theta) = \begin{pmatrix}
			\cosh^2\theta & \sinh^2\theta & -\frac{1}{2} \sinh2\theta \\
			\sinh^2\theta & \cosh^2\theta & - \frac{1}{2} \sinh2\theta \\
			-\sinh2\theta & -\sinh2\theta & \cosh2\theta
		\end{pmatrix}, \quad
		\det M(\theta) = 1.
		\label{eq:supp_Mtheta}
	\end{equation}
	Its pair-changing block is
	\begin{equation}
		A(\theta) = \begin{pmatrix}
			\cosh^2\theta & \sinh^2\theta \\
			\sinh^2\theta & \cosh^2\theta
		\end{pmatrix}, \quad
		\det A(\theta) = \cosh2\theta.
		\label{eq:supp_A_theta}
	\end{equation}
	Thus the factor $\cosh 2\theta$ in Eq.~\eqref{eq:supp_u2w2_det} arises from the pair-changing subblock of the full canonical transformation.
	
	For a quadratic channel,
	\begin{equation}
		R = A_0 a^2 + B_0 a^{\dagger2},
	\end{equation}
	the squeezing transformation gives
	\begin{equation}
		R \to q_1 a^2 + q_2 a^{\dagger2} + q_0 (2a^\dagger a + 1),
		\label{eq:supp_q_channels}
	\end{equation}
	where
	\begin{equation}
		q_1 = A_0 \cosh^2\theta + B_0 \sinh^2\theta, \quad
		q_2 = A_0 \sinh^2\theta + B_0\cosh^2\theta, \quad
		q_0 = - \frac{1}{2} (A_0+B_0)\sinh2\theta.
		\label{eq:supp_q_coefficients}
	\end{equation}
	The Gaussian triangularization imposes $q_1=0$, from which
	\begin{equation}
		\frac{q_0}{q_2} = -\frac{1}{2} \tanh2\theta.
		\label{eq:supp_q_ratio}
	\end{equation}
	The residual $iJ_y$ channel therefore takes the form
	\begin{equation}
		\frac{w_2}{\sqrt{N}} \left[ a^{\dagger2} - \left(a^\dagger a + \frac{1}{2} \right) \tanh2\theta \right] iJ_y,
		\label{eq:supp_residual_two_photon}
	\end{equation}
	consistent with Eq.~\eqref{eq:supp_H2s}.
	
	Projection onto the Gaussian vacuum eliminates the $a^{\dagger 2}$ term and retains the number-preserving contribution. Using $iJ_y=(J_+-J_-)/2$, the interaction-generated $J_+$ coefficient in the highest-weight boundary equation is
	\begin{equation}
		\Sigma_{\mathrm{edge}}^{(2)} = - \frac{w_2}{4 \sqrt{N}} \tanh2\theta.
		\label{eq:supp_Sigma2_first}
	\end{equation}
	Since
	\begin{equation}
		\frac{\tanh2\theta}{\sqrt{N}} = \frac{u_2}{\omega \cosh2\theta},
	\end{equation}
	Eqs.~\eqref{eq:supp_u2w2_det} and \eqref{eq:supp_Sigma2_first} give
	\begin{equation}
		\Sigma_{\mathrm{edge}}^{(2)} = - \frac{u_2w_2}{4\omega \cosh2\theta} =
		\frac{\det G}{2\omega}.
		\label{eq:supp_Sigma2_det}
	\end{equation}
	Thus the additional Bogoliubov factor in the pair-changing block is compensated by the number-preserving contribution.
	
	The canonical matrix counterpart of this reciprocal factor follows from the full Bogoliubov block structure. Writing
	\begin{equation}
		M = \begin{pmatrix}
			A&b \\ c^T&d
		\end{pmatrix},
	\end{equation}
	the Schur complement of the pair-changing block satisfies
	\begin{equation}
		d - c^T A^{-1} b = \frac{\det M}{\det A} = (\cosh2\theta)^{-1}.
		\label{eq:supp_Schur_D2}
	\end{equation}
	Since $\det M(\theta)=1$ whereas $\det A(\theta) = \cosh 2\theta$, the Schur complement necessarily contributes the reciprocal factor $(\cosh 2\theta)^{-1}$. This is precisely the factor entering Eq.~\eqref{eq:supp_Sigma2_det}, which restores the determinant dependence of the full highest-weight boundary coefficient.
	
	\subsection{Determinant form of the Juddian condition}
	\label{sec:supp_invariant_Juddian}
	
	The interaction-generated highest-weight boundary coefficients obtained from the one- and two-photon constructions coincide,
	\begin{equation}
		\Sigma_{\mathrm{edge}}^{(1)} = \Sigma_{\mathrm{edge}}^{(2)} = \frac{\det G}{2\omega}.
		\label{eq:supp_Sigma_common}
	\end{equation}
	The highest-weight boundary condition therefore becomes
	\begin{equation}
		\left( -\frac{\Delta}{2} + \frac{\det G}{2\omega} \right) \bra{j,j} J_+ \ket{j,j-1} c_{0,j-1}^{(k)} = 0.
		\label{eq:supp_invariant_boundary}
	\end{equation}
	For the lowest Juddian branch constructed in Sec.~\ref{sec:supp_exact_closure}, $c_{0,j-1}^{(k)} \neq 0$, while $\bra{j,j} J_+ \ket{j,j-1} = \sqrt{2j} \neq 0$. Consequently,
	\begin{equation}
		\det G(g_J) = \Delta\omega.
		\label{eq:supp_detG_Juddian}
	\end{equation}
	Using Eq.~\eqref{eq:supp_detG}, $\det G=g^2(1-r^2)$, this gives
	\begin{equation}
		g_J^2 = \frac{\Delta\omega}{1-r^2}.
		\label{eq:supp_gJ_det}
	\end{equation}
	
	Equations~\eqref{eq:supp_Sigma_common}-\eqref{eq:supp_detG_Juddian} complete the determinant derivation: both Gaussian constructions yield the same interaction-generated highest-weight boundary coefficient, $\det G/(2\omega)$, and hence the same Juddian condition.
	
	\subsection{Channel-resolved normal-state instabilities}
	\label{sec:supp_gaussian_curvature}
	
	We finally determine the channel-resolved normal-state instability thresholds. Introduce the generalized bosonic quadratures
	\begin{equation}
		X_k = \frac{a^k+a^{\dagger k}}{\sqrt{N}}, \quad
		Y_k = i \frac{a^k-a^{\dagger k}}{\sqrt{N}},
		\label{eq:supp_XP_channels}
	\end{equation}
	in terms of which the interaction is
	\begin{equation}
		H_{\mathrm{int}}^{(k)} = G_+ X_k J_x + G_- Y_k J_y, \quad 
		G_\pm = g (1 \pm r).
		\label{eq:supp_Hint_XP}
	\end{equation}
	
	To determine the local curvature about the normal state, we treat the collective spin semiclassically while retaining the bosonic mode at the Gaussian level. Define
	\begin{equation}
		m_x = \frac{\braket{J_x}}{j}, \quad
		m_y = \frac{\braket{J_y}}{j}, \quad
		j=\frac{N}{2}.
	\end{equation}
	Near the normal state, $\braket{J_z} / j \approx -1$, and
	\begin{equation}
		\frac{\braket{J_z}}{j} = - \sqrt{1-m_x^2-m_y^2} = -1 + \frac{m_x^2+m_y^2}{2} + O(m_\perp^4), \quad 
		m_\perp^2 = m_x^2+m_y^2.
		\label{eq:supp_Jz_normal_expansion}
	\end{equation}
	The spin contribution is therefore
	\begin{equation}
		E_{\mathrm{s}} = - \frac{N\Delta}{2} + \frac{N\Delta}{4} \left(m_x^2 + m_y^2 \right) + O(m_\perp^4).
		\label{eq:supp_Espin_curvature}
	\end{equation}

	For fixed $(m_x, m_y)$, the bosonic Hamiltonian is
	\begin{equation}
		H_{\mathrm{b}}^{(k)} = \omega a^\dagger a + \frac{N}{2} G_+ m_x X_k + \frac{N}{2} G_- m_y Y_k.
		\label{eq:supp_Hboson_channels}
	\end{equation}
	For $k=1$, displacement gives the ground-state energy shift
	\begin{equation}
		\delta E_{\mathrm{b}}^{(1)} = - \frac{N}{4\omega} \left( G_+^2 m_x^2 + G_-^2 m_y^2 \right).
		\label{eq:supp_Eb1_channels}
	\end{equation}
	
	For $k=2$, 
	\begin{equation}
		H_{\mathrm{b}}^{(2)} = \omega a^\dagger a +
		\xi a^2 + \xi^*a^{\dagger2},
		\quad
		\xi= \frac{\sqrt{N}}{2} G_+ m_x + i \frac{\sqrt{N}}{2} G_- m_y.
		\label{eq:supp_Hb2_channels}
	\end{equation}
	Its Gaussian excitation frequency is $\sqrt{ \omega^2 - 4|\xi|^2}$, and the corresponding ground-state energy shift is
	\begin{equation}
		\delta E_{\mathrm{b}}^{(2)} = \frac{1}{2} \left( \sqrt{\omega^2 - 4|\xi|^2} - \omega \right) = - \frac{|\xi|^2}{\omega} + O(|\xi|^4) = - \frac{N}{4\omega} \left( G_+^2 m_x^2 + G_-^2 m_y^2 \right) + O(m_\perp^4).
		\label{eq:supp_Eb2_channels}
	\end{equation}
	Thus the one- and two-photon models have the same quadratic  normal-state curvature.
	
	Combining the spin and bosonic contributions gives the energy density expanded about the normal state
	\begin{equation}
		\frac{E_{\mathrm{loc}}^{(k)}}{N} = - \frac{\Delta}{2} + \frac{1}{4} \left( \Delta - \frac{G_+^2}{\omega} \right) m_x^2 + \frac{1}{4} \left( \Delta - \frac{G_-^2}{\omega} \right) m_y^2 + O(m_\perp^4).
		\label{eq:supp_MF_curvature}
	\end{equation}
	The two quadratic coefficients in Eq.~\eqref{eq:supp_MF_curvature} vanish separately when
	\begin{equation}
		G_\pm^2 = \Delta\omega.
		\label{eq:supp_channel_instability}
	\end{equation}
	Accordingly, the channel-resolved normal-state instability thresholds are
	\begin{equation}
		g_\pm = \frac{\sqrt{\Delta \omega}}{1 \pm r}, \quad
		g_+<g_-,
		\label{eq:supp_lambda_pm}
	\end{equation}
	where the ordering follows from $0<r<1$.
	
	For the one-photon ADM, $g_+$ becomes the thermodynamic superradiant critical coupling. For the two-photon ADM, $g_\pm$ characterize local instabilities of the normal-state curvature and are not identified here with thermodynamic phase boundaries.
	
	Their product is
	\begin{equation}
		g_+ g_- = \frac{\Delta\omega}{1-r^2}.
		\label{eq:supp_lambda_product}
	\end{equation}
	Combining Eq.~\eqref{eq:supp_lambda_product} with the exact Juddian condition~\eqref{eq:supp_gJ_det} yields
	\begin{equation}
		g_J^2 = g_+ g_-, \quad
		g_J = \sqrt{g_+ g_-}, \quad 
		g_+<g_J<g_-.
		\label{eq:supp_geometric_mean}
	\end{equation}
	Thus the individual channel strengths $G_\pm$ set the normal-state instability thresholds, whereas their product $G_+G_- = \det G$ fixes the exact Juddian crossing.
	
	\section{PARITY-RESOLVED JUDDIAN DOUBLETS AND FINITE-$N$ CUSPS}
	\label{sec:supp_parity}
	
	Starting from the exact Gaussian product eigenstates constructed in Sec.~\ref{sec:supp_truncation}, we form the parity-resolved Juddian doublets, establish their ground-state character, and prove that their crossing has a nonzero slope discontinuity. The two-channel representation of the crossing condition was established independently in Sec.~\ref{sec:supp_determinant}.
	
	\subsection{Parity-resolved Juddian doublets}
	\label{sec:supp_parity_branches}
	
	The two Hamiltonians conserve the generalized parity
	\begin{equation}
		\Pi_k = \exp \! \left[ i\pi\left( \frac{a^\dagger a}{k} + J_z + j \right)\right], \quad
		[\Pi_k, H_k] = 0.
		\label{eq:supp_Pik}
	\end{equation}
	For $k=1$, $\Pi_1^2 = 1$. For $k=2$, $\Pi_2^4=1$, with eigenvalues $\pm 1$ and $\pm i$; within the even-boson subspace containing the lowest squeezed Juddian branch, only $\Pi_2=\pm1$ occur.
	
	Let $\ket{\Psi_L^{(k)}}$ denote the normalized Gaussian product eigenstate constructed in Sec.~\ref{sec:supp_truncation}. For the
	collective-spin coherent states,
	\begin{equation}
		e^{i \pi (J_z + j)} \ket{\zeta}_j = (-1)^N \ket{-\zeta}_j, \quad 2j=N.
		\label{eq:supp_spin_zeta_flip}
	\end{equation}
	The corresponding bosonic transformations are
	\begin{equation}
		e^{i\pi a^\dagger a} \ket{-\alpha} = \ket{\alpha}, \quad
		e^{i\pi a^\dagger a/2} S(-\theta) e^{-i\pi a^\dagger a/2} = S(\theta).
		\label{eq:supp_boson_parity_action}
	\end{equation}
	The Juddian branches are therefore
	\begin{align}
		&\ket{\Psi_L^{(1)}} = \ket{-\alpha} \otimes \ket{\zeta_1}_j, \quad
		\ket{\Psi_R^{(1)}} = \ket{\alpha} \otimes \ket{-\zeta_1}_j, \label{eq:supp_branches1}\\
		&\ket{\Psi_L^{(2)}} = \ket{-\theta}_2 \otimes \ket{\zeta_2}_j, \quad
		\ket{\Psi_R^{(2)}} = \ket{\theta}_2 \otimes \ket{-\zeta_2}_j.
		\label{eq:supp_branches2}
	\end{align}
	Since $[\Pi_k,H_k]=0$, the two branches are exact
	eigenstates with the same Juddian energy, and
	\begin{equation}
		\Pi_k \ket{\Psi_L^{(k)}} = (-1)^N \ket{\Psi_R^{(k)}}, \quad
		\Pi_k \ket{\Psi_R^{(k)}} = (-1)^N \ket{\Psi_L^{(k)}}.
		\label{eq:supp_branch_parity}
	\end{equation}

	Define the interbranch overlap as
	\begin{equation}
		S_k \equiv \braket{\Psi_L^{(k)}|\Psi_R^{(k)}}.
		\label{eq:supp_Sk_def}
	\end{equation}
	The normalized parity-resolved Juddian states are
	\begin{equation}
		\ket{\Psi_{J,\pm}^{(k)}} = \frac{ \ket{\Psi_L^{(k)}} \pm (-1)^N \ket{\Psi_R^{(k)}}}{ \sqrt{2(1 \pm (-1)^N S_k)}}.
		\label{eq:supp_general_cats}
	\end{equation}
	They satisfy
	\begin{equation}
		\Pi_k\ket{\Psi_{J,\pm}^{(k)}} = \pm \ket{\Psi_{J, \pm}^{(k)}}, \quad
		H_k \ket{\Psi_{J,\pm}^{(k)}} = E_J^{(k)} \ket{\Psi_{J,\pm}^{(k)}}.
		\label{eq:supp_cat_properties}
	\end{equation}
	
	For the one-photon branches,
	\begin{equation}
		S_1 = \braket{-\alpha|\alpha} {}_j \! \braket{\zeta_1|-\zeta_1}_j = e^{-2\alpha^2} (-\eta_1)^N,
		\label{eq:supp_S1}
	\end{equation}
	where 
	\begin{equation}
		{}_j \! \braket{\zeta|-\zeta}_j = \left( \frac{1-\zeta^2}{1+\zeta^2} \right)^N, \quad
		\frac{1-\zeta_1^2}{1+\zeta_1^2} = -\eta_1 = -\rho.
		\label{eq:supp_spin_overlap1}
	\end{equation}
	For the two-photon branches, the bosonic interbranch overlap is
	\begin{equation}
		B_2 \equiv {}_2 \! \braket{-\theta|\theta}_2 = \frac{1}{\sqrt{\cosh2\theta}}, 
		\label{eq:supp_B2}
	\end{equation}
	while
	\begin{equation}
		\frac{1-\zeta_2^2}{1+\zeta_2^2} = -\eta_2.
		\label{eq:supp_spin_overlap2}
	\end{equation}
	Hence
	\begin{equation}
		S_2 = (-\eta_2)^N B_2 = \frac{(-\eta_2)^N}{\sqrt{\cosh2\theta}}.
		\label{eq:supp_S2}
	\end{equation}
	
	For $0<r<1$, $0<\eta_1<1$. Throughout the normalizable two-photon Juddian regime, $0<\eta_2<1$ and $\cosh2\theta \geqslant 1$. Consequently,
	\begin{equation}
		|S_1|<1, \quad
		|S_2|<1, \quad
		\operatorname{sgn}(S_k) = (-1)^N.
		\label{eq:supp_overlap_properties}
	\end{equation}
	Since $1\pm(-1)^NS_k>0$, both states in Eq.~\eqref{eq:supp_general_cats} are well defined and nonzero. They form an exactly degenerate opposite-parity Juddian doublet at $g=g_J$ in either model.
	
	\subsection{Ground-state character of the Juddian doublet}
	\label{sec:supp_ground_state_proof}
	
	We establish the ground-state character of the parity-resolved Juddian states by constructing exact positive-semidefinite factorizations of the full Hamiltonians at $g=g_J$.
	
	For the one-photon model, define 
	\begin{equation}
		Q_n^{(1)} = \sqrt{\frac{\omega}{N}} a + \frac{g_J}{\sqrt \omega} \left( \sigma_-^{(n)} + r \sigma_+^{(n)} \right), \quad 
		n=1,\ldots,N.
		\label{eq:supp_Q1_psd}
	\end{equation}
	Using
	\begin{equation}
		\left( \sigma_+ + r\sigma_- \right) \left( \sigma_- + r\sigma_+ \right) = \frac{1+r^2}{2} + \frac{1-r^2}{2} \sigma_z
	\end{equation}
	together with $g_J^2 (1-r^2) = \Delta\omega$, direct expansion gives 
	\begin{equation}
		H_1(g_J) - E_J^{(1)} = \sum_{n=1}^{N} Q_n^{(1)\dagger} Q_n^{(1)}
		\geqslant 0.
		\label{eq:supp_psd_one_photon}
	\end{equation}
	
	For the two-photon ADM, introduce the matrices
	\begin{equation}
		U =
		\begin{pmatrix}
			\sqrt{\frac{\omega}{N} + \frac{E_J^{(2)}}{N} - \frac{\Delta}{2}} & 0 \\
			0 & \sqrt{\frac{\omega}{N} + \frac{E_J^{(2)}}{N} + \frac{\Delta}{2}}
		\end{pmatrix},
		\quad
		V =
		\begin{pmatrix}
			0 & \sqrt{- \frac{E_J^{(2)}}{N} - \frac{\Delta}{2}} \\
			\sqrt{- \frac{E_J^{(2)}}{N} + \frac{\Delta}{2}} & 0
		\end{pmatrix}.
		\label{eq:supp_UV_psd}
	\end{equation}
	Substituting Eq.~\eqref{eq:supp_EJ2} and the Juddian condition into Eq.~\eqref{eq:supp_UV_psd} shows that all four radicands are nonnegative for $0<\Delta<\Delta_c$. The matrices $U$ and $V$ therefore satisfy
	\begin{align}
		U^\dagger U + V^\dagger V =
		\frac{\omega}{N} , \quad
		V^\dagger V =  - \frac{E_J^{(2)}}{N} + \frac{\Delta}{2} \sigma_z, \quad
		U^\dagger V = \frac{g_J}{\sqrt{N}} \left( r\sigma_+ + \sigma_- \right), \quad
		V^\dagger U = \frac{g_J}{\sqrt{N}} \left( \sigma_+ + r\sigma_- \right).
		\label{eq:supp_UV_identities}
	\end{align}
	For each spin $n$, define
	\begin{equation}
		Q_n^{(2)} = U^{(n)} a + V^{(n)}a^\dagger,
		\label{eq:supp_Q2_psd}
	\end{equation}
	where $U^{(n)}$ and $V^{(n)}$ act only on the $n$th spin. Since the bosonic operators commute with the spin operators, Eqs.~\eqref{eq:supp_UV_identities} give
	\begin{equation}
		H_2(g_J)-E_J^{(2)} = \sum_{n=1}^{N} Q_n^{(2)\dagger} Q_n^{(2)}
		\geqslant 0.
		\label{eq:supp_psd_two_photon}
	\end{equation}
	
	Equations~\eqref{eq:supp_psd_one_photon} and \eqref{eq:supp_psd_two_photon} establish $E_J^{(k)}$ as a global spectral lower bound through exact operator identities on the full spin-boson Hilbert space. The two nonzero parity-resolved eigenstates constructed in Sec.~\ref{sec:supp_parity_branches} both saturate this bound and are therefore exactly degenerate ground states.
	
	\subsection{Hellmann-Feynman identity and finite-$N$ cusp}
	\label{sec:supp_HF_identity}
	
	In the $(g,\Delta)$ plane, the Juddian condition defines the curve
	\begin{equation}
		\Delta_J(g) = \frac{(1-r^2)g^2}{\omega}.
		\label{eq:supp_DeltaJ}
	\end{equation}
	Along the physical Juddian line, the two parity-resolved branches are exactly degenerate,
	\begin{equation}
		E_{J,+}^{(k)} \! \left[g, \Delta_J(g)\right] = E_{J,-}^{(k)} \! \left[g, \Delta_J(g)\right] \equiv E_J^{(k)}(g).
		\label{eq:supp_degenerate_manifold}
	\end{equation}
	Applying the Hellmann-Feynman theorem separately in the two parity sectors gives
	\begin{equation}
		\frac{dE_J^{(k)}}{dg} = \bra{\Psi_{J,\pm}^{(k)}} \partial_g H_k \ket{\Psi_{J,\pm}^{(k)}} + \frac{d\Delta_J}{dg} \braket{J_z}_\pm,
		\label{eq:supp_HF_each}
	\end{equation}
	where $\partial_g$ denotes differentiation with respect to $g$ at fixed $\Delta$, and
	\begin{equation}
		\braket{J_z}_\pm \equiv \bra{\Psi_{J,\pm}^{(k)}} J_z \ket{\Psi_{J,\pm}^{(k)}}.
	\end{equation}
	Since
	\begin{equation}
		\frac{d\Delta_J}{dg} = \frac{2(1-r^2)g}{\omega},
	\end{equation}
	subtracting the two parity-sector versions of Eq.~\eqref{eq:supp_HF_each} gives the magnitude of the fixed-$\Delta$ ground-state slope discontinuity, which defines the cusp strength,
	\begin{equation}
		\mathcal{J}_X^{(k)} \equiv \left| \bra{\Psi_{J,+}^{(k)}} \partial_g H_k \ket{\Psi_{J,+}^{(k)}} - \bra{\Psi_{J,-}^{(k)}} \partial_g H_k \ket{\Psi_{J,-}^{(k)}} \right| = \frac{2 (1-r^2) g_J}{\omega} \left| \braket{J_z}_+ - \braket{J_z}_- \right|.
		\label{eq:supp_JX_Jz}
	\end{equation}

	To evaluate the remaining matrix elements, write
	\begin{equation}
		\ket{\Psi_{J,\pm}^{(k)}} = \frac{ \ket{b_L^{(k)}} \ket{\zeta_k}_j \pm (-1)^N \ket{b_R^{(k)}} \ket{-\zeta_k}_j }{ \sqrt{2(1 \pm (-1)^N S_k)}},
	\end{equation}
	and define the bosonic interbranch overlap
	\begin{equation}
		B_k \equiv \braket{b_L^{(k)}|b_R^{(k)}}.
	\end{equation}
	For real $\zeta_k$,
	\begin{equation}
		{}_j \! \braket{\zeta_k|-\zeta_k}_j = (-\eta_k)^N, \quad
		{}_j \! \bra{\zeta_k} J_z \ket{\zeta_k}_j = - \frac{N}{2} \eta_k, \quad
		{}_j \! \bra{\zeta_k} J_z \ket{-\zeta_k}_j = \frac{N}{2} (-\eta_k)^{N-1},
	\end{equation}
	Using $S_k = B_k (-\eta_k)^N$, one obtains
	\begin{equation}
		\braket{J_z}_+ - \braket{J_z}_- = - N B_k \frac{ \eta_k^{N-1} (1-\eta_k^2) }{ 1-S_k^2 }.
		\label{eq:supp_Jz_difference_general}
	\end{equation}
	
	For the one-photon ADM,
	\begin{equation}
		B_1 = e^{-2\alpha^2},
	\end{equation}
	and therefore
	\begin{equation}
		\left| \braket{J_z}_+ - \braket{J_z}_- \right| = Ne^{-2\alpha^2} \frac{ \eta_1^{N-1} (1-\eta_1^2) }{ 1-S_1^2}.
		\label{eq:supp_Jz_difference1}
	\end{equation}
	Using
	\begin{equation}
		\frac{1-\eta_1^2}{\eta_1} = \frac{4r}{1-r^2}, \quad 
		|S_1| = e^{-2\alpha^2}\eta_1^N,
	\end{equation}
	Eq.~\eqref{eq:supp_JX_Jz} becomes
	\begin{equation}
		\mathcal{J}_X^{(1)} = \frac{8Nrg_J}{\omega} \frac{|S_1|}{1-S_1^2}.
		\label{eq:supp_JX1_exact}
	\end{equation}
	
	For the two-photon ADM,
	\begin{equation}
		B_2 = \frac{1}{\sqrt{\cosh2\theta}} = \sqrt{\frac{\beta_+}{\beta_-}},
	\end{equation}
	consistent with Eq.~\eqref{eq:supp_B2}. The Juddian condition also gives 
	\begin{equation}
		\frac{(1-r^2)g_J}{\omega} = \frac{\Delta}{g_J},
	\end{equation}
	Hence
	\begin{equation}
		\mathcal{J}_X^{(2)} = \frac{2N\Delta}{g_J} \sqrt{\frac{\beta_+}{\beta_-}} \frac{ \eta_2^{N-1} (1-\eta_2^2)}{ 1-S_2^2 }.
		\label{eq:supp_JX2_exact}
	\end{equation}
	
	For $0<r<1$, one has $0<\eta_1<1$. Throughout the normalizable two-photon Juddian regime $g_J<g_c$, $0<\eta_2<1$, $\beta_\pm>0$, and $|S_2|<1$; moreover, $|S_1|<1$ for the one-photon branch. Consequently,
	\begin{equation}
		\mathcal{J}_X^{(1)}>0, \quad
		\mathcal{J}_X^{(2)}>0
		\label{eq:supp_JX_positive}
	\end{equation}
	for every finite $N$ and every regular point on the corresponding physical Juddian branch.
	
	To establish the first-order character, consider the two parity-resolved energy branches at fixed $\Delta$ near $g=g_J$. Their difference has the local expansion
	\begin{equation}
		E_+^{(k)}(g)-E_-^{(k)}(g) = \left[ \partial_g E_+^{(k)} - \partial_g E_-^{(k)} \right]_{g=g_J} (g-g_J) + O \! \left( (g-g_J)^2 \right).
		\label{eq:supp_local_branch_crossing}
	\end{equation}
	By Eq.~\eqref{eq:supp_JX_positive}, the linear coefficient is nonzero, and the two opposite-parity branches exchange their energy ordering across $g_J$. 
	
	Since Sec.~\ref{sec:supp_ground_state_proof} establishes that the degenerate Juddian states saturate the global ground-state lower bound at the crossing, the derivative of the ground-state energy has a jump of magnitude
	\begin{equation}
		\left| \left. \partial_g E_{\mathrm{GS}}^{(k)} \right|_{g_J^+} - \left. \partial_g E_{\mathrm{GS}}^{(k)} \right|_{g_J^-} \right| = \mathcal{J}_X^{(k)} > 0.
		\label{eq:supp_EGS_derivative_jump}
	\end{equation}
	Thus the symmetry-protected Juddian ground-state crossing is an exact finite-$N$ first-order transition at every regular physical point of the Juddian branch.
	
	\subsection{Coherent suppression in the one-photon model}
	\label{sec:supp_coherent_suppression}
	
	The one-photon cusp strength in Eq.~\eqref{eq:supp_JX1_exact} is controlled by the interbranch overlap $S_1$. At the Juddian crossing,
	\begin{equation}
		\alpha^2 = \frac{Nrg_J^2}{\omega^2} = \frac{Nr}{1-r^2}\frac{\Delta}{\omega},
		\label{eq:supp_alpha_J}
	\end{equation}
	and hence
	\begin{equation}
		|S_1| = \left[ \eta_1 \exp \! \left( - \frac{2r}{1-r^2} \frac{\Delta}{\omega} \right)\right]^N.
		\label{eq:supp_S1_N}
	\end{equation}
	Using the channel-resolved normal-state instability thresholds defined in Eq.~\eqref{eq:supp_lambda_pm},
	\begin{equation}
		\eta_1 = \frac{1-r}{1+r} = \frac{g_+}{g_-} = \left( \frac{g_+}{g_J} \right)^2, \quad g_J^2 = g_+g_-.
		\label{eq:supp_eta1_lambda}
	\end{equation}
	Equation~\eqref{eq:supp_S1_N} becomes
	\begin{equation}
		|S_1| = \left[ \frac{1-r}{1+r} \exp \! \left( - \frac{2r}{1-r^2} \frac{\Delta}{\omega} \right) \right]^N = \left[ \exp \! \left( - \frac{2r}{1-r^2} \frac{\Delta}{\omega} \right) \left( \frac{g_+}{g_J} \right)^2 \right]^N.
		\label{eq:supp_S1_lambda}
	\end{equation}
	
	For fixed $r$ and $\Delta/\omega$, define
	\begin{equation}
		c \equiv \frac{2r}{1-r^2} \frac{\Delta}{\omega} - 2 \ln \! \left(\frac{g_+}{g_J}\right) = \frac{2r}{1-r^2} \frac{\Delta}{\omega} + \ln \! \left( \frac{1+r}{1-r} \right).
		\label{eq:supp_c_one_photon}
	\end{equation}
	For $0<r<1$ and $\Delta/\omega>0$, one has $c>0$, so
	\begin{equation}
		|S_1| = e^{-cN}.
	\end{equation}
	Since $S_1^2=O(e^{-2cN})$, Eq.~\eqref{eq:supp_JX1_exact} gives
	\begin{equation}
		\mathcal J_X^{(1)} = \frac{8rg_J}{\omega} N e^{-cN} \left[ 1 + O \! \left( e^{-2cN} \right) \right], \quad
		N \to \infty.
		\label{eq:supp_JX1_exp}
	\end{equation}
	Thus the one-photon ground-state cusp is exponentially suppressed in the thermodynamic limit.
	
	A distinct asymptotic suppression occurs at fixed finite $N$ as $\Delta / \omega \to \infty$. Equation~\eqref{eq:supp_S1_N} gives
	\begin{equation}
		|S_1| = \eta_1^N \exp \! \left[ - \frac{2Nr}{1-r^2} \frac{\Delta}{\omega} \right].
	\end{equation}
	Together with Eq.~\eqref{eq:supp_JX1_exact}, one obtains
	\begin{equation}
		\mathcal J_X^{(1)} = O \! \left[ \sqrt{ \frac{\Delta}{\omega} } \exp \! \left( - \frac{2Nr}{1-r^2} \frac{\Delta}{\omega} \right) \right]
		\to 0, \quad
		\frac{\Delta}{\omega} \to \infty
	\end{equation}
	at fixed finite $N$.
	
	Equation~\eqref{eq:supp_JX_positive} guarantees $\mathcal J_X^{(1)}>0$ at every regular finite point on the one-photon Juddian line. The cusp strength therefore vanishes only in the thermodynamic or fixed-$N$, infinite frequency-ratio limit.
	
	\section{SPECTRAL-COLLAPSE TERMINATION AND ENDPOINT SCALING}
	\label{sec:supp_distinct_fates}
	
	In the two-photon model, the algebraic Juddian condition yields a discrete eigenstate only when the corresponding squeezed state is normalizable. We determine this boundary and analyze the asymptotic finite-$N$ scaling laws near the resulting endpoint.
	
	\subsection{Normalizability and the collapse boundary}
	\label{sec:supp_two_photon_admissibility}
	
	For the two-photon Juddian branch,
	\begin{equation}
		\beta_\pm^2 = 1-\frac{Ng_J^2}{\omega^2} (1 \pm r)^2.
		\label{eq:supp_beta_J_repeat}
	\end{equation}
	For $0<r<1$, $\beta_+^2$ reaches zero before $\beta_-^2$ as $g_J$ is increased. The normalizability condition $\beta_+>0$ therefore gives
	\begin{equation}
		g_J<g_c, \quad
		g_c = \frac{\omega}{\sqrt{N} (1+r)},
		\label{eq:supp_gc}
	\end{equation}
	Thus $g_c$ is the boundary of the normalizable two-photon Juddian branch.
	
	The same scale is exposed by the loss of bosonic confinement in the extremal-$J_x$ channel. In the spin states $\ket{j,\pm j}_x$, one has
	\begin{equation}
		{}_x \! \bra{j,\pm j} J_y \ket{j,\pm j}_x = {}_x \! \bra{j,\pm j} J_z \ket{j,\pm j}_x = 0.
		\label{eq:supp_extremal_spin_expectations}
	\end{equation}
	Projecting the full two-photon Hamiltonian onto the extremal spin states $\ket{j,\pm j}_x$ gives the bosonic quadratic forms
	\begin{equation}
		H_{{\rm b},\pm} = \omega a^\dagger a \pm \frac{\sqrt N}{2} G_+(g) \left( a^2 + a^{\dagger 2} \right).
		\label{eq:supp_extremal_bosonic}
	\end{equation}
	Both quadratic forms have the Bogoliubov frequency
	\begin{equation}
		\Omega_{\rm b} = \sqrt{\omega^2-NG_+^2(g)}.
	\end{equation}
	At
	\begin{equation}
		\sqrt N G_+(g_c) = \omega,
		\label{eq:supp_collapse_condition}
	\end{equation}
	the quadratic confinement vanishes. For $g>g_c$, the quadratic form in Eq.~\eqref{eq:supp_extremal_bosonic} has an unstable
	direction, so bosonic trial states can lower its expectation value without bound. Equation~\eqref{eq:supp_collapse_condition} reproduces the normalizability threshold $g_c$ in Eq.~\eqref{eq:supp_gc}.
	
	Combining Eq.~\eqref{eq:supp_gc} with the exact Juddian condition defines the endpoint value
	\begin{equation}
		\Delta_c \equiv \Delta_J (g_c) = \frac{\omega}{N} \frac{1-r}{1+r} = \frac{\rho \omega}{N}.
		\label{eq:supp_Deltac}
	\end{equation}
	Equivalently,
	\begin{equation}
		\frac{g_J^2}{g_c^2} = \frac{\Delta}{\Delta_c},
		\label{eq:supp_endpoint_ratio}
	\end{equation}
	and hence
	\begin{equation}
		g_J < g_c \quad \Longleftrightarrow \quad
		\Delta < \Delta_c.
		\label{eq:supp_admissible_equivalence}
	\end{equation}
	
	Approaching the limiting point
	\begin{equation}
		(g_J,\Delta) \to (g_c,\Delta_c)
		\label{eq:supp_endpoint}
	\end{equation}
	from the normalizable side, 
	\begin{equation}
		\beta_+ \to 0^+, \quad 
		\cosh2\theta = \frac{\beta_-}{\beta_+} \to \infty.
	\end{equation}
	The squeezed Juddian state therefore loses normalizability at the collapse boundary. For $\Delta>\Delta_c$, the algebraic Juddian condition continues to define $g_J$, but the corresponding state is nonnormalizable and no longer represents a physical discrete eigenstate. The physical finite-$N$ FOT line therefore terminates at the spectral-collapse endpoint $(g_c,\Delta_c)$.
	
	\subsection{Finite-$N$ FOT termination scaling}
	\label{sec:supp_two_photon_termination}
	
	Along the normalizable Juddian FOT line, define the dimensionless distance to the collapse boundary by
	\begin{equation}
		s \equiv \beta_+^2 = 1 - \frac{g_J^2}{g_c^2} = 1 - \frac{\Delta}{\Delta_c}, \quad
		s \to 0^+.
		\label{eq:supp_s_line}
	\end{equation}
	Using $\rho = (1-r)/(1+r)$, one obtains
	\begin{equation}
		\beta_-^2 = 1 - \rho^2  + \rho^2 s, \quad
		\beta_- = \sqrt{1-\rho^2} [1+O(s)].
		\label{eq:supp_beta_minus_endpoint}
	\end{equation}
	Hence
	\begin{equation}
		\eta_2 = \rho \frac{\beta_+}{\beta_-} = \frac{\rho}{\sqrt{1 - \rho^2}} s^{1/2} [1+O(s)],
		\label{eq:supp_eta2_scaling}
	\end{equation}
	while the bosonic interbranch overlap defined in Eq.~\eqref{eq:supp_B2} satisfies
	\begin{equation}
		B_2 = \sqrt{\frac{\beta_+}{\beta_-}} = (1 - \rho^2)^{-1/4} s^{1/4} [1+O(s)].
		\label{eq:supp_B2_endpoint}
	\end{equation}
	The total interbranch overlap therefore scales as
	\begin{equation}
		|S_2| = B_2\eta_2^N =  \rho^N (1-\rho^2)^{-\frac{2N+1}{4}} s^{\frac{2N+1}{4}} [1+O(s)].
		\label{eq:supp_S2_endpoint}
	\end{equation}
	
	The exact finite-$N$ cusp strength derived in Eq.~\eqref{eq:supp_JX2_exact} is
	\begin{equation}
		\mathcal{J}_X^{(2)} = \frac{2N\Delta}{g_J} \sqrt{\frac{\beta_+}{\beta_-}} \frac{ \eta_2^{N-1} (1-\eta_2^2) }{ 1-S_2^2 }.
		\label{eq:supp_JX2_endpoint_start}
	\end{equation}
	As $s \to 0^+$,
	\begin{equation}
		1-\eta_2^2 = 1+O(s), \quad
		S_2^2 =
		O \! \left( s^{\frac{2N+1}{2}} \right).
	\end{equation}
	The overlap-dependent factor in Eq.~\eqref{eq:supp_JX2_endpoint_start} then becomes
	\begin{equation}
		\sqrt{\frac{\beta_+}{\beta_-}} \eta_2^{N-1} = \rho^{N-1} \left( \frac{\beta_+}{\beta_-} \right)^{N-\frac12} =
		\rho^{N-1} (1-\rho^2)^{\frac14-\frac N2} s^{(2N-1)/4} [1+O(s)].
	\end{equation}
	The remaining prefactor is regular at the endpoint:
	\begin{equation}
		\frac{2N\Delta}{g_J} = \frac{2N \Delta_c}{g_c} [1+O(s)] = 2 \sqrt N (1-r) [1+O(s)].
		\label{eq:supp_cusp_prefactor_endpoint}
	\end{equation}
	Combining Eqs.~\eqref{eq:supp_JX2_endpoint_start} - \eqref{eq:supp_cusp_prefactor_endpoint} gives
	\begin{equation}
		\mathcal{J}_X^{(2)} = A_N(r) s^{\kappa_N} \left[ 1+O(s) \right], \quad
		\kappa_N = \frac{2N-1}{4},
		\label{eq:supp_JX2_scaling}
	\end{equation}
	with
	\begin{equation}
		A_N(r) = 2 \sqrt{N} (1-r) \rho^{N-1} \left( 1-\rho^2 \right)^{\frac14 - \frac N2}.
		\label{eq:supp_AN}
	\end{equation}
	For every finite $N$ and $0<r<1$, $A_N(r)$ is finite and strictly positive. Thus $\kappa_N=(2N-1)/4$ is the termination exponent governing the vanishing of the ground-state cusp strength along the physical Juddian FOT line.
	
	Finally,
	\begin{equation}
		s = 1 - \frac{g_J^2}{g_c^2} = \frac{2(g_c-g_J)}{g_c} + O \!\left[ (g_c-g_J)^2 \right], \quad
		g_J \to g_c^-,
	\end{equation}
	so that
	\begin{equation}
		\mathcal{J}_X^{(2)} \propto |g_J-g_c|^{\kappa_N}, \quad
		\kappa_N = \frac{2N-1}{4}.
		\label{eq:supp_JX2_gc_scaling}
	\end{equation}
	The finite-$N$ ground-state cusp therefore tends continuously to zero as the physical Juddian FOT line approaches the collapse endpoint.
	
	\subsection{Branch-resolved endpoint theory and soft-mode scaling}
	\label{sec:supp_gaussian_endpoint}
	
	To analyze the endpoint away from the Juddian line, define
	\begin{equation}
		s \equiv 1-\frac{g^2}{g_c^2}, \quad
		s \to 0^+ .
		\label{eq:supp_s_off}
	\end{equation}
	The collapse boundary is $s=0$. Along the Juddian line, Eq.~\eqref{eq:supp_s_off} reduces to $s = \beta_+^2 = 1 - g_J^2/g_c^2$.
	
	We first identify the spin backgrounds selected by the exact two-photon Juddian states at the endpoint. As $g_J \to g_c^-$, $\beta_+ \to 0$ and $\cosh 2\theta = \beta_-/\beta_+ \to \infty$, so that
	\begin{equation}
		\zeta_2^2 = \frac{(1+r) \cosh2\theta + (1-r)}{(1+r) \cosh2\theta - (1-r)} \to 1 .
		\label{eq:supp_zeta2_to_one}
	\end{equation}
	For real collective-spin coherent states $\ket{\pm \zeta_2}_j$,
	\begin{equation}
		\frac{\braket{J_x}}{j} = \pm \frac{2\zeta_2}{1+\zeta_2^2}, \quad \frac{\braket{J_z}}{j} = \frac{1-\zeta_2^2}{1+\zeta_2^2}, \quad
		\braket{J_y}=0,
		\label{eq:supp_spin_coherent_expectation}
	\end{equation}
	and hence
	\begin{equation}
		\ket{\zeta_2}_j \to \ket{j,+j}_x, \quad
		\ket{-\zeta_2}_j \to \ket{j,-j}_x.
		\label{eq:supp_Jx_endpoint}
	\end{equation}
	We therefore use the extremal configurations $J_x=\pm j$ as the reference spin states for the endpoint expansion.
	
	We approach the endpoint along the fixed-$\Delta_c$ trajectory, $\Delta=\Delta_c$ and $g\to g_c^-$. Introducing 
	\begin{equation}
		x=\frac{a+a^\dagger}{\sqrt2}, \quad
		p=\frac{a-a^\dagger}{i\sqrt2},
		\label{eq:supp_quadratures}
	\end{equation}
	and
	\begin{equation}
		g_c = \frac{\omega}{\sqrt N(1+r)}, \quad
		\Delta_c = \frac{\omega \rho}{N}, \quad
		\rho = \frac{1-r}{1+r},
	\end{equation}
	the two-photon Hamiltonian becomes
	\begin{align}
		H_2 =& \frac{\omega}{2} (x^2+p^2-1) + \Delta_c J_z + \frac{\omega g}{Ng_c} (x^2-p^2) J_x - \omega \frac{g\rho}{Ng_c} (xp+px) J_y \nonumber\\
		=& \frac{\omega}{2} \left[ \left( 1+\frac{2gJ_x}{Ng_c} \right) x^2 + \left( 1-\frac{2gJ_x}{Ng_c} \right) p^2 - 1 \right] + \Delta_c J_z - \Delta_c \frac{g}{g_c} (xp+px) J_y .
		\label{eq:supp_H2_quadrature}
	\end{align}
	
	The diagonal bosonic Hamiltonians in the two extremal spin backgrounds are
	\begin{align}
		H_{00}^{(-)} &= \frac{\omega}{2} \left[ \left( 1 - \sqrt{1-s} \right)x^2 + \left( 1 + \sqrt{1-s} \right) p^2 - 1 \right] \approx \omega \left[ \frac{s}{4} x^2 + \left( 1 - \frac{s}{4} \right) p^2 - \frac12 \right],
		\label{eq:supp_projected_minus} \\
		H_{00}^{(+)} &= \frac{\omega}{2} \left[ \left( 1 + \sqrt{1-s} \right)x^2 + \left( 1 - \sqrt{1-s} \right) p^2 - 1 \right] \approx \omega \left[ \left( 1 - \frac{s}{4} \right)x^2 + \frac{s}{4} p^2 - \frac12 \right].
		\label{eq:supp_projected_plus}
	\end{align}
	Thus the soft coordinates are $x$ for the $J_x=-j$ branch and $p$ for the $J_x=+j$ branch.
	
	The diagonal projections are insufficient because $J_y$ and $J_z$ couple the extremal spin states to neighboring $J_x$ sectors. We therefore eliminate the leading contributions from these neighboring sectors to obtain the effective branch-local Hamiltonians.
	
	\paragraph{Soft branch around $J_x=-j$.}
	Write
	\begin{equation}
		\ket{\Psi_-} = \psi_0^{(-)}(x) \ket{j,-j}_x + \psi_1^{(-)}(x) \ket{j,-j+1}_x + \cdots.
		\label{eq:supp_xbasis_expansion_minus}
	\end{equation}
	Choose the phase of $\ket{j,-j+1}_x$ such that
	\begin{equation}
		{}_x \!\bra{j,-j} J_y \ket{j,-j+1}_x = \frac{\sqrt N}{2}, \quad
		{}_x \! \bra{j,-j} J_z \ket{j,-j+1}_x = \frac{i\sqrt N}{2}.
		\label{eq:supp_xbasis_matrix_minus}
	\end{equation}
	At $(g_c,\Delta_c)$, the first two projected equations are
	\begin{align}
		\left( E + \frac{\omega}{2} \right)\psi_0^{(-)} = & - \omega \partial_x^2 \psi_0^{(-)} + i \frac{\omega \rho}{\sqrt N} (1 + x\partial_x) \psi_1^{(-)},
		\label{eq:supp_soft_eq0_minus} \\
		\left( E + \frac{\omega}{2} \right) \psi_1^{(-)} = & \frac{\omega}{N} x^2 \psi_1^{(-)} - \omega \frac{N-1}{N} \partial_x^2 \psi_1^{(-)} + i \frac{\omega \rho}{\sqrt N} x \partial_x \psi_0^{(-)} + i \frac{\omega \rho \sqrt{2(N-1)}}{N} (1 + x\partial_x) \psi_2^{(-)} .
		\label{eq:supp_soft_eq1_minus}
	\end{align}
	At the endpoint, the $J_x=-j$ channel has no $x^2$ confinement, so the low-energy wave function extends to asymptotically large $|x|$. In this regime, the diagonal $x^2\psi_1^{(-)}/N$ term controls the leading elimination,
	\begin{equation}
		\psi_1^{(-)}(x) \approx -i \rho \sqrt N \frac{\partial_x \psi_0^{(-)}(x)}{x}.
		\label{eq:supp_psi1_elimination_minus}
	\end{equation}
	Using
	\begin{equation}
		(1+x\partial_x) \frac{\partial_x \psi_0^{(-)}}{x} = \partial_x^2 \psi_0^{(-)},
	\end{equation}
	one obtains
	\begin{equation}
		\left( E + \frac{\omega}{2} \right) \psi_0^{(-)} \approx -\omega (1-\rho^2) \partial_x^2 \psi_0^{(-)}.
		\label{eq:supp_endpoint_kinetic_minus}
	\end{equation}
	
	\paragraph{Soft branch around $J_x=+j$.}
	In the momentum representation, write
	\begin{equation}
		\ket{\Psi_+} = \psi_0^{(+)}(p) \ket{j,+j}_x + \psi_1^{(+)}(p) \ket{j,+j-1}_x + \cdots.
		\label{eq:supp_xbasis_expansion_plus}
	\end{equation}
	Choose the phase of $\ket{j,+j-1}_x$ such that
	\begin{equation}
		{}_x \! \bra{j,+j} J_y \ket{j,+j-1}_x = \frac{\sqrt N}{2}, \quad
		{}_x \! \bra{j,+j} J_z \ket{j,+j-1}_x = - \frac{i\sqrt N}{2}.
		\label{eq:supp_xbasis_matrix_plus}
	\end{equation}
	Using $x=i\partial_p$, the first two endpoint equations are
	\begin{align}
		\left( E + \frac{\omega}{2} \right) \psi_0^{(+)} =&
		-\omega \partial_p^2 \psi_0^{(+)} - i \frac{\omega \rho}{\sqrt N} (1 + p\partial_p) \psi_1^{(+)}, \label{eq:supp_soft_eq0_plus} \\
		\left( E + \frac{\omega}{2} \right) \psi_1^{(+)} =&
		\frac{\omega}{N} p^2 \psi_1^{(+)} - \omega \frac{N-1}{N} \partial_p^2 \psi_1^{(+)} - i \frac{\omega \rho}{\sqrt N} p \partial_p \psi_0^{(+)} - i \frac{\omega \rho \sqrt{2(N-1)}}{N} (1+p\partial_p) \psi_2^{(+)} .
		\label{eq:supp_soft_eq1_plus}
	\end{align}
	The leading-order elimination yields
	\begin{equation}
		\psi_1^{(+)}(p) \approx i \rho \sqrt N \frac{\partial_p \psi_0^{(+)}(p)}{p},
		\label{eq:supp_psi1_elimination_plus}
	\end{equation}
	and therefore
	\begin{equation}
		\left( E + \frac{\omega}{2} \right) \psi_0^{(+)} \approx - \omega (1-\rho^2) \partial_p^2 \psi_0^{(+)} .
		\label{eq:supp_endpoint_kinetic_plus}
	\end{equation}
	
	For $g<g_c$ sufficiently close to the endpoint, the coefficients of the projected equations vary regularly with $g/g_c = \sqrt{1-s}$. The kinetic coefficient therefore remains finite, $1 - \rho^2 + O(s)$, whereas the restoring curvature of the corresponding soft coordinate is $s/4+O(s^2)$. To leading order, 
	\begin{align}
		H_{\mathrm{eff}}^{(-)} + \frac{\omega}{2} &\approx \omega \left[ - (1-\rho^2) \partial_x^2 + \frac{s}{4} x^2 \right],
		\label{eq:supp_Heff_minus} \\
		H_{\mathrm{eff}}^{(+)} + \frac{\omega}{2} &\approx \omega \left[ - (1-\rho^2) \partial_p^2 + \frac{s}{4} p^2 \right].
		\label{eq:supp_Heff_plus}
	\end{align}
	The characteristic widths of the branch-local soft-mode wave functions scale as
	\begin{equation}
		w_x \propto s^{-1/4} \quad (J_x=-j), \quad
		w_p \propto s^{-1/4} \quad (J_x=+j),
		\label{eq:supp_soft_coordinate_scaling}
	\end{equation}
	Accordingly, derivatives acting on the soft-mode envelopes contribute factors of order $s^{1/4}$. Within this power counting, the derivative, energy, and higher-spin-sector terms omitted in the leading elimination are subleading as $s \to 0^+$. The nonextremal spin components are estimated to satisfy $\psi_\ell^{(\pm)} = O(s^{\ell/2}) \psi_0^{(\pm)}$.
	
	The two branch-local effective Hamiltonians are related by the two-photon parity transformation
	\begin{equation}
		\Pi_2: \quad J_x \to -J_x, \quad
		x \to p, \quad p \to -x,
		\label{eq:supp_parity_branch_map}
	\end{equation}
	and are therefore isospectral. Their common adjacent branch-local oscillator spacing is
	\begin{equation}
		\Omega_{\mathrm{eff}} = \omega \sqrt{1-\rho^2} s^{1/2} [1+O(s)].
		\label{eq:supp_Omega_eff}
	\end{equation}
	As a consistency check, along the Juddian line
	\begin{align}
		\Omega_J \equiv \omega \beta_+\beta_- = \omega \sqrt{ s \left( 1 - \rho^2 + \rho^2 s \right)} = \omega \sqrt{ 1 - \rho^2} s^{1/2} [1+O(s)].
		\label{eq:supp_Omega_match}
	\end{align}
	
	Since the lowest two-photon Juddian branch is restricted to the even-boson subspace, the branch-local harmonic-oscillator eigenstates in this even-boson subspace have oscillator quantum numbers $n=0,2,4,\ldots$. Therefore the lowest allowed soft excitation within a fixed parity sector is separated from the ground state by two adjacent oscillator spacings,
	\begin{equation}
		\epsilon_{\rm sp} = 2 \Omega_{\mathrm{eff}} \propto s^{1/2} \propto |g-g_c|^{1/2}, \quad
		g \to g_c^-.
		\label{eq:supp_epssp}
	\end{equation}
	Thus the same-parity soft-mode exponent is $1/2$, independent of $N$.
	
	The branch-local theory does not determine the parametrically smaller splitting between opposite-parity combinations of the two branches. That different-parity gap is analyzed separately in the following subsection.

	\subsection{Endpoint scaling of the different-parity gap}
	\label{sec:supp_two_direction}
	
	At a regular point of the Juddian line, the ground-state slope discontinuity determines the leading splitting between the two parity-resolved branches. Defining
	\begin{equation}
		\epsilon_{\mathrm{dp}}^{(k)} \equiv \left|E_+^{(k)}-E_-^{(k)}\right|,
	\end{equation}
	the fixed-$\Delta$ expansion about $g=g_J$ gives
	\begin{equation}
		\epsilon_{\mathrm{dp}}^{(k)} = \mathcal{J}_X^{(k)} |g-g_J| + O \! \left[ (g-g_J)^2 \right].
		\label{eq:supp_regular_dp_splitting}
	\end{equation}
	In the two-photon ADM, however, $\mathcal J_X^{(2)}$ itself vanishes as the Juddian line approaches the collapse endpoint.  The joint endpoint limit therefore requires an assessment of the higher-order detuning terms.
	
	We approach the endpoint along the fixed-$\Delta_c$ trajectory. For fixed $g<g_c$, write
	\begin{equation}
		H_2 (g, \Delta) = H_2 \! \left[g, \Delta_J(g) \right] + \delta \Delta J_z, \quad
		\delta \Delta \equiv \Delta - \Delta_J(g).
		\label{eq:supp_H_off_Juddian}
	\end{equation}
	Because $[J_z,\Pi_2] = 0$, the detuning perturbation preserves parity. For fixed $s>0$, Rayleigh-Schr\"odinger perturbation theory can therefore be performed independently in the two parity sectors. To first order,
	\begin{equation}
		\epsilon_{\mathrm{dp}} = |\delta \Delta| \left| \braket{J_z}_+ - \braket{J_z}_- \right|_J + O(\delta \Delta^2),
		\label{eq:supp_epsdp_first}
	\end{equation}
	where the expectation values are evaluated in the exact parity-resolved Juddian states.
	
	On the two-photon Juddian line,
	\begin{equation}
		\left| \braket{J_z}_+ - \braket{J_z}_- \right|_J = N \sqrt{ \frac{\beta_+}{\beta_-}} \frac{ \eta_2^{N-1} (1-\eta_2^2) }{ 1-S_2^2 } = N \rho^{N-1} \left( 1 - \rho^2 \right)^{ \frac14 - \frac N2} s^{\kappa_N} \left[ 1+O(s) \right], \quad
		\kappa_N = \frac{2N-1}{4}.
		\label{eq:supp_Jz_endpoint}
	\end{equation}
	Along $\Delta=\Delta_c$,
	\begin{equation}
		\delta\Delta = \Delta_c-\Delta_J(g) = \Delta_c \left( 1 - \frac{g^2}{g_c^2} \right) = \Delta_c s.
		\label{eq:supp_t_critical}
	\end{equation}
	The first-order contribution therefore scales as $s^{\kappa_N+1}$. We next estimate the endpoint scaling of higher-order parity-sensitive corrections using spin-tail suppression, interbranch Gaussian overlaps, and the same-parity soft-mode energy scale.
	
	For fixed $s>0$, expand 
	\begin{equation}
		E_\pm (s, \delta \Delta) = E_J(s) + \sum_{m=1}^{\infty} (\delta \Delta)^m \mathcal E_\pm^{(m)}(s).
		\label{eq:supp_energy_expansion}
	\end{equation}
	As the endpoint is approached, the two Juddian branches tend to the extremal spin configurations $m_x=\pm j$, which are separated by $2j=N$ spin steps. In the $J_x$ basis,
	\begin{equation}
		{}_x \! \bra{m_x'} J_z \ket{m_x}_x \neq 0
		\quad \Longrightarrow \quad
		m_x' = m_x \pm 1 .
		\label{eq:supp_Jz_selection}
	\end{equation}
	Hence each insertion of $J_z$ changes $m_x$ by at most one unit. The nonextremal spin tails of the exact Juddian states are controlled by
	\begin{equation}
		\eta_2 = \rho \frac{\beta_+}{\beta_-} =
		O(s^{1/2}),
		\label{eq:supp_eta2_endpoint}
	\end{equation}
	so a component $n$ spin steps away from either extremal configuration has amplitude $O(\eta_2^n)$.
	
	Only interbranch contributions can distinguish the two
	parity-resolved combinations. For an $m$th-order process with $1 \leqslant m \leqslant N$, the $m$ explicit insertions of $J_z$ can supply at most $m$ of the $N$ spin steps. In the leading spin-tail estimate, the remaining $N-m$ steps are assigned the suppression factor
	\begin{equation}
		O \! \left(\eta_2^{N-m}\right).
		\label{eq:supp_residual_spin_bound}
	\end{equation}
	
	A parity-sensitive contribution must also connect the two bosonic squeezing backgrounds. For squeezed oscillator states $\ket{\nu;\theta} = S(\theta)\ket{\nu}$ with fixed quantum numbers $\nu$ and $\nu'$, the interbranch matrix element can be written as
	\begin{equation}
		{}_2 \! \braket{\nu';-\theta|\nu;\theta}_2 = B_2 f_{\nu \nu'}(\tanh 2\theta), \quad
		B_2 = \frac{1}{\sqrt{\cosh2\theta}} = \sqrt{\frac{\beta_+}{\beta_-}}.
		\label{eq:supp_gaussian_cross_overlap}
	\end{equation}
	For fixed $\nu$ and $\nu'$, $f_{\nu \nu'}$ remains bounded as $s \to 0^+$. The corresponding interbranch matrix element is therefore $O(B_2) = O(s^{1/4})$. In the following power counting, we assume that this overlap suppression remains effective in the relevant intermediate-state contributions.
	
	The remaining possible infrared enhancement arises from intermediate-state reduced resolvents. Because the perturbation preserves parity, the opposite-parity Juddian partner does not enter the reduced resolvent. For the low-energy reduced resolvent, we use the estimate
	\begin{equation}
		R_\pm = O(\Omega_J^{-1}) =O(s^{-1/2}),
		\label{eq:estimate_resolvent}
	\end{equation}
	assuming that no softer same-parity excitation scale contributes to the intermediate-state sums. The following order-by-order power counting uses this resolvent scale together with the spin-tail and interbranch-overlap estimates above. An $m$th-order coefficient contains at most $m-1$ such soft resolvents. Hence, for $1 \leqslant m \leqslant N$, 
	\begin{equation}
		\mathcal E_+^{(m)} - \mathcal E_-^{(m)} = O \! \left[ B_2 \eta_2^{N-m} \Omega_J^{-(m-1)} \right].
		\label{eq:supp_mth_coefficient}
	\end{equation}
	
	Along $\Delta = \Delta_c$,
	\begin{equation}
		\frac{|\delta \Delta|}{\epsilon_{\rm sp}} \propto s^{1/2} \to 0.
	\end{equation}
	At fixed finite $N$, $\|J_z\| \leqslant N/2$, so
	$\| \delta\Delta J_z \| / \epsilon_{\rm sp} = O(s^{1/2}) \to 0$. Thus the detuning is perturbatively small within each parity sector. The endpoint scaling of the parity-sensitive terms additionally depends on the overlap and resolvent estimates above.
	
	Using
	\begin{equation}
		B_2 \propto s^{1/4}, \quad
		\eta_2 \propto s^{1/2}, \quad
		\Omega_J \propto s^{1/2}, \quad
		\delta \Delta \propto s,
	\end{equation}
	we obtain, for $1 \leqslant m \leqslant N$,
	\begin{equation}
		(\delta \Delta)^m \left( \mathcal E_+^{(m)} - \mathcal E_-^{(m)} \right) = O \! \left[ B_2 \eta_2^{N-m} \frac{ (\delta \Delta)^m }{ \Omega_J^{m-1} } \right]
		= O \! \left[ s^{1/4} s^{(N-m)/2} s^m s^{-(m-1)/2} \right]
		= O \! \left( s^{\frac{2N+3}{4}} \right).
		\label{eq:supp_mth_scaling}
	\end{equation}
	Under the spin-tail, overlap, and resolvent estimates specified above, the power counting gives contributions of order $O(s^{\mu_N})$ for $1 \leqslant m \leqslant N$, where $\mu_N = (2N+3) / 4$. Contributions from different orders may occur at the same power and modify the leading coefficient.
	
	For $m>N$, the perturbative sequence can supply all $N$ spin steps, so no residual spin-tail factor is required. Parity sensitivity still requires at least one interbranch bosonic overlap, giving
	\begin{equation}
		(\delta \Delta)^m \left( \mathcal E_+^{(m)} - \mathcal E_-^{(m)} \right) = O \! \left[ B_2 \frac{ (\delta \Delta)^m }{ \Omega_J^{m-1} } \right] = O \! \left( s^{\frac{m}{2} + \frac34} \right), \quad
		m>N.
		\label{eq:supp_higher_m_scaling}
	\end{equation}
	For $m>N$, we have
	\begin{equation}
		\frac{m}{2}+\frac34 > \frac{N}{2}+\frac34 =
		\frac{2N+3}{4}.
	\end{equation}
	Thus, at each fixed order $m>N$, the estimated contribution is subleading to $s^{\mu_N}$.
	
	The low-energy power counting suggests that higher-order parity-sensitive contributions introduce no lower power of $s$, although terms of orders $2 \leqslant m \leqslant N$ may modify the coefficient of $s^{\mu_N}$. Together with the exact nonzero first-order coefficient, this motivates the following prediction for the full different-parity gap, consistent with the numerical results in Fig.~3(a) of the main text:
	\begin{equation}
		\epsilon_{\mathrm{dp}} \propto s^{\mu_N} \propto |g-g_c|^{\mu_N}, \quad
		\mu_N = \kappa_N+1 = \frac{2N+3}{4}.
		\label{eq:supp_muN}
	\end{equation}
	For $N=1$, $\mu_1=5/4$, recovering the single-qubit result of Ref.~\cite{liQuantumCriticalitySpectral2026}. For $N>1$, $\mu_N$ acquires an explicit $N$ dependence inherited from the finite-$N$ FOT cusp exponent.
	
	\subsection{Quantum Fisher information}
	\label{sec:supp_qfi}
	
	Near the collapse endpoint, the soft-mode results of Sec.~\ref{sec:supp_gaussian_endpoint} and the predicted parity-gap scaling of Sec.~\ref{sec:supp_two_direction} give
	\begin{equation}
		\epsilon_{\mathrm{sp}} \propto s^{1/2}, \quad
		\epsilon_{\mathrm{dp}} \propto s^{\mu_N}, \quad
		\mu_N = \frac{2N+3}{4} > \frac12.
		\label{eq:supp_two_gaps}
	\end{equation}
	Hence $\epsilon_{\mathrm{dp}} \ll \epsilon_{\mathrm{sp}}$ as $ g \to g_c^-$. We now determine which sector contributes to the singular ground-state response.
	
	Along the fixed-$\Delta_c$ trajectory with $g<g_c$, the quantum Fisher information (QFI) of the nondegenerate ground state $\ket{\Psi_0(g)}$ with respect to $g$ is
	\begin{equation}
		F_Q = 4 \left[ \braket{\partial_g \Psi_0 | \partial_g \Psi_0} - \left| \braket{\Psi_0 | \partial_g\Psi_0} \right|^2 \right] = 4 \sum_{n\neq0} \frac{ \left| \bra{\Psi_n} \partial_g H_2 \ket{\Psi_0} \right|^2 }{ (E_n-E_0)^2 }.
		\label{eq:supp_FQ}
	\end{equation}
	
	Since $\Pi_2$ is independent of $g$ and
	\begin{equation}
		[\Pi_2, H_2(g)]=0, 
	\end{equation}
	one has
	\begin{equation}
		[\Pi_2, \partial_gH_2]=0.
	\end{equation}
	Therefore,
	\begin{equation}
		\bra{\Psi_{n,\mp}} \partial_g H_2 \ket{\Psi_{0,\pm}} = 0.
		\label{eq:supp_QFI_opposite_zero}
	\end{equation}
	The different-parity excitation therefore makes no contribution to the QFI spectral sum, independently of the magnitude of its gap.
	
	The leading singular contribution follows from the branch-local
	effective Hamiltonians in Eqs.~\eqref{eq:supp_Heff_minus} and \eqref{eq:supp_Heff_plus}. Writing either branch in terms of its soft coordinate $q=x$ or $p$,
	\begin{equation}
		H_{\mathrm{eff}} + \frac{\omega}{2} = \omega \left[ - A(s) \partial_q^2 + B(s) q^2 \right],
		\label{eq:supp_QFI_Heff}
	\end{equation}
	with
	\begin{equation}
		A(s) = 1 - \rho^2 + O(s), \quad
		B(s) = \frac{s}{4} + O(s^2).
		\label{eq:supp_QFI_AB}
	\end{equation}
	The normalized branch-local ground-state envelope is
	\begin{equation}
		\psi_0 (q;s) = \left( \frac{\lambda(s)}{\pi} \right)^{1/4} \exp\! \left[ - \frac{\lambda(s)}{2} q^2 \right],
		\label{eq:supp_QFI_gaussian}
	\end{equation}
	with
	\begin{equation}
		\lambda(s) = \sqrt{ \frac{B(s)}{A(s)} } = \frac{ s^{1/2} }{ 2 \sqrt{1-\rho^2} } \left[ 1+O(s) \right].
		\label{eq:supp_QFI_lambda}
	\end{equation}
	
	For the normalized real Gaussian state in Eq.~\eqref{eq:supp_QFI_gaussian}, direct evaluation gives
	\begin{equation}
		F_Q^{\mathrm{soft}} = \frac12 \left( \frac{\mathrm{d} \ln\lambda}{\mathrm{d} g} \right)^2.
		\label{eq:supp_QFI_gaussian_metric}
	\end{equation}
	Using
	\begin{equation}
		s = 1 - \frac{g^2}{g_c^2}, \quad
		\frac{ds}{dg} = - \frac{2g}{g_c^2},
	\end{equation}
	one obtains
	\begin{equation}
		\frac{\mathrm{d} \ln\lambda}{dg} = \frac{1}{2s} \frac{ds}{dg} + O(1) = - \frac{g}{g_c^2s} + O(1),
		\label{eq:supp_QFI_dlambda}
	\end{equation}
	and hence
	\begin{equation}
		F_Q^{\mathrm{soft}} = \frac{g^2}{2g_c^4} s^{-2} \left[ 1+O(s) \right] = \frac{1}{2g_c^2} s^{-2} \left[ 1+O(s) \right].
		\label{eq:supp_FQ_soft}
	\end{equation}
	Thus the branch-local soft mode produces an $s^{-2}$ singularity.
	
	The virtual spin admixture entering the endpoint construction
	gives only a weaker singularity. For example, Eq.~\eqref{eq:supp_psi1_elimination_minus} yields
	\begin{equation}
		\psi_1^{(-)} \approx - i \rho \sqrt N \frac{ \partial_x\psi_0^{(-)} }{x}.
	\end{equation}
	For the Gaussian envelope,
	\begin{equation}
		\frac{ \partial_x \psi_0^{(-)} }{x} = - \lambda(s) \psi_0^{(-)},
	\end{equation}
	and therefore
	\begin{equation}
		\frac{ \psi_1^{(-)} }{ \psi_0^{(-)} } = O \! \left( s^{1/2} \right).
	\end{equation}
	The same scaling holds for both symmetry-related soft branches. Differentiation with respect to $g$ then gives at most
	\begin{equation}
		F_Q^{\mathrm{spin}} = O \! \left( s^{-1} \right),
		\label{eq:supp_FQ_spin}
	\end{equation}
	which is subleading to the Gaussian soft-mode contribution. Higher nonextremal spin components are more strongly suppressed.

	The two branch-local Hamiltonians have the same Gaussian metric by parity symmetry. Combining the selection rule \eqref{eq:supp_QFI_opposite_zero} with the asymptotic estimates \eqref{eq:supp_FQ_soft} and \eqref{eq:supp_FQ_spin} gives
	\begin{equation}
		F_Q \propto s^{-2}.
		\label{eq:supp_FQ_s}
	\end{equation}
	Since
	\begin{equation}
		s = 1-\frac{g^2}{g_c^2} = \frac{2(g_c-g)}{g_c} + O \! \left[ (g_c-g)^2 \right], \quad
		g\to g_c^-,
	\end{equation}
	the QFI obeys
	\begin{equation}
		F_Q \propto |g-g_c|^{-2}.
		\label{eq:supp_FQ_gc}
	\end{equation}
	The QFI exponent is therefore independent of $N$ and is governed by the same-parity branch-local soft mode, despite the parametrically smaller different-parity gap.

	
	%